\documentclass[fleqn,usenatbib]{mnras}

\usepackage{newtxtext,newtxmath}

\usepackage[T1]{fontenc}
\usepackage{ae,aecompl}
\pdfoutput=1

\usepackage{graphicx}	
\usepackage{amsmath}	
\usepackage{amssymb}	
\usepackage{enumitem}

\usepackage[usenames, dvipsnames]{color}
\definecolor{mygray}{gray}{0.4}

\title[Mass and environment quenching in EAGLE]{Dissecting the roles of mass and environment quenching in galaxy evolution with EAGLE}

\author[Cochrane \& Best]{
R. K. Cochrane$^{1,2}$\thanks{E-mail: rcoch@roe.ac.uk} \& P. N. Best$^{1}$
\\
$^{1}$SUPA, Institute for Astronomy, Royal Observatory Edinburgh, EH9 3HJ, UK\\
$^{2}$Isaac Newton Group of Telescopes, E-38700 Santa Cruz de La Palma, Canary Islands, Spain
}

\date{Accepted 2018 June 26. Received 2018 June 26; in original form 2018 February 19}

\pubyear{2018}

\volume{{\rm in press}}

\begin{document}
\label{firstpage}
\pagerange{\pageref{firstpage}--\pageref{lastpage}}
\maketitle

\begin{abstract}\\
We exploit the pioneering cosmological hydrodynamical simulation, EAGLE, to study how the connection between halo mass ($M_{\rm halo}$), stellar mass ($M_*$) and star-formation rate (SFR) evolves across redshift. Using Principal Component Analysis we identify the key axes of correlation between these physical quantities, for the full galaxy sample and split by satellite/central and low/high halo mass. The first principal component of the $z=0$ EAGLE galaxy population is a positive correlation between $M_{\rm halo}$, $M_*$ and SFR. This component is particularly dominant for central galaxies in low mass haloes. The second principal component, most significant in high mass haloes, is a negative correlation between $M_{\rm halo}$ and SFR, indicative of environmental quenching. For galaxies above $M_{*}\sim10^{10}M_{\odot}$, however, the SFR is seen to decouple from the $M_{\rm halo}$--$M_{*}$ correlation; this result is found to be independent of environment, suggesting that mass quenching effects are also in operation. We find extremely good agreement between the EAGLE principal components and those of SDSS galaxies; this lends confidence to our conclusions. Extending our study to EAGLE galaxies in the range $z=0-4$, we find that, although the relative numbers of galaxies in the different subsamples change, their principal components do not change significantly with redshift. This indicates that the physical processes that govern the evolution of galaxies within their dark matter haloes act similarly throughout cosmic time. Finally, we present halo occupation distribution model fits to EAGLE galaxies and show that one flexible 6-parameter functional form is capable of fitting a wide range of different mass- and SFR-selected subsamples.
\end{abstract}


\section{Introduction}
In most theories of galaxy formation and evolution, halo mass is a key ingredient. It is generally accepted that galaxies form and grow under the gravity of dark matter haloes \citep{White1978}, which themselves form via successive mergers and accretion events. This process happens within the large-scale structure of sheets, filaments and the nodes where they intersect, together known as the `cosmic web' \citep{Bond1996}. The spatial distribution of galaxies reflects this, with the largest clusters of galaxies residing in the densest dark matter overdensities. \\
\indent It has long been known that there are clear differences between the physical properties of galaxies in different environments; in particular field galaxies are more likely to be star-forming and morphologically disk-like than those in the more overdense regions of rich groups and clusters, at low and moderate redshifts \citep[e.g.][]{Oemler1977,Dressler1980,Balogh2000,Sobral2011,Boselli2016,Kelkar2017}. The most massive haloes host massive galaxies that assembled stars earlier \citep[e.g.][]{Tojeiro2017}, and are less efficient at forming stars at low redshift. However, the extent to which these trends are driven by local density as compared to the direct influence of the cosmic web remains unclear \citep[e.g.][]{Eardley2015}. \\
\indent A myriad of recent work in extragalactic astrophysics has focused on revealing the physical processes which drive galaxy `quenching', the process by which a previously star-forming galaxy halts star formation and becomes passive. \cite{Peng2010} suggests that these could be separated into two separate (and independent) quenching modes: `mass quenching' (most high-mass galaxies are passive) and `environment quenching' (most galaxies in clusters are passive, regardless of their mass). The latter has been proposed to be primarily important for satellite galaxies, with the satellite quenching process being more closely linked to local galaxy density than overall halo mass \citep{Peng2012}. However, others interpret the same data as indicating a stronger role of halo mass. \cite{Woo2013} show that the passive fraction of central galaxies is more correlated with halo mass at fixed stellar mass than with stellar mass at fixed halo mass. For satellite galaxies, there is a strong dependence on both halo mass and distance to the halo centre. \cite{Woo2013} suggest that local overdensity measurements can be unreliable and dependent on the number of observed group members, and instead argue that the halo mass is the key driver of quenching. \\
\indent \cite{Gabor2014} argue that both mass and environment quenching can be attributed to hot gas in massive host dark matter haloes \citep[see also][]{Birnboim2013,Keres2005}. Below some characteristic dark matter halo mass (typically $\sim10^{12}M_{\odot}$, the approximate peak of the stellar mass - halo mass relation, \citealt{Moster2010}), gas cooling times are short compared to the dynamical time of the dark matter halo, and cold gas accretes efficiently and forms stars \citep{Dekel2006}. Above this halo mass, cooling times are long, and the gas that accretes onto the galaxy is hot, so star-formation is inefficient. \cite{Bower2016} explore this in more detail, proposing that the effectiveness of star-formation-driven outflows depends on their buoyancy compared to that of the halo. Above some characteristic halo mass scale, these outflows are unable to clear gas from the galaxy, resulting in the buildup of gas in the central regions which then drives a rapid increase in black hole mass. This, in turn, heats the halo, preventing further gas accretion. Galaxies are then not replenished with fuel for star formation, and star formation in high mass haloes is thus inefficient \citep[see also][for observational evidence for quenching via gas-exhaustion, or `strangulation']{Peng2015}. Similar arguments have been made within radio AGN feedback models, whereby the presence of hot intracluster gas in more massive dark matter haloes provides both a fuel source and an energy repository for recurrent radio AGN activity, which acts as a self-regulating feedback cycle controlling gas cooling rates and hence star formation \citep[e.g. see the review by][]{Heckman2014}.\\
\indent Investigating whether two physically distinct quenching mechanisms are really required by the data, \cite{Zu2015} study whether quenching is primarily driven by stellar mass or halo mass by modelling the clustering and weak lensing of galaxies in SDSS. They conclude that models in which the quenching of both central and satellite galaxies depends solely on halo mass (but in different ways) provide the best fit to observations, without the need for a second variable such as galaxy stellar mass. Furthermore, they find a critical quenching mass of $M_{\rm{halo}}\sim1.5\times10^{12}h^{-1}M_{\odot}$ for both central and satellite galaxies. \\
\indent Despite this work, the influence of the dark matter halo on its galaxies is not understood in detail. This is partly due to the inherent difficulties of linking galaxies to their host haloes observationally. Normally, this is attempted using one of two methods: Halo Occupation distribution (HOD; \citealt{Ma2000,Peacock2000}, see \citealt{Cooray2002} for a review) modelling, whereby the occupation of haloes as a function of mass is modelled for central and satellite galaxies separately, then fitted to clustering or weak-lensing observations; and Subhalo Abundance Matching (SHAM; \citealt{Conroy2006}), which traditionally assigns galaxies to dark matter haloes by ranking them by stellar mass and subhalo mass (e.g. as measured by circular velocity). This becomes more difficult when we seek to explore different populations of galaxies (i.e. in those selected in terms of mass, star-formation rate or colour). \\
\indent In this paper, we take a simpler approach. We draw simulated galaxies and their host haloes directly from the Virgo Consortium's Evolution and Assembly of GaLaxies and their Environments project, known as EAGLE \citep{Crain2015,Schaye2015,McAlpine2015}. EAGLE is state-of-the-art in cosmological hydrodynamical simulations. By tuning subgrid models of feedback from massive stars and AGN to the observed low-redshift galaxy stellar mass function, galaxy size-galaxy mass relation and galaxy mass - black hole mass relation, EAGLE has been able to match observations on which it has not been calibrated (e.g. galaxy specific star-formation rate distributions, passive fractions and the Tully-Fisher relation, \citealt{Schaye2015}) far better than past hydrodymical simulations. \\
\indent In Section \ref{sec:hm_sm_sfr}, we introduce the sample and present the relationships between stellar mass, halo mass and star-formation rate as seen by EAGLE over cosmic time. In Section \ref{sec:pca_overall}, we quantify the strength of these relations using a statistical technique, Principal Component Analysis, over the redshift range $z=0-4$. We also compare the $z=0$ results to observational data from SDSS using an equivalent analysis. We discuss the implications of our results for the quenching of star-formation in Section \ref{sec:quenching_modes}, and draw conclusions in Section \ref{sec:conclusions}. An appendix to the paper explores the halo occupation of galaxies in different stellar mass and star-formation rate bins in EAGLE, as this is a key input to studies that use HOD fitting.
\begin{figure*} 
	\centering
	\includegraphics[scale=0.45]{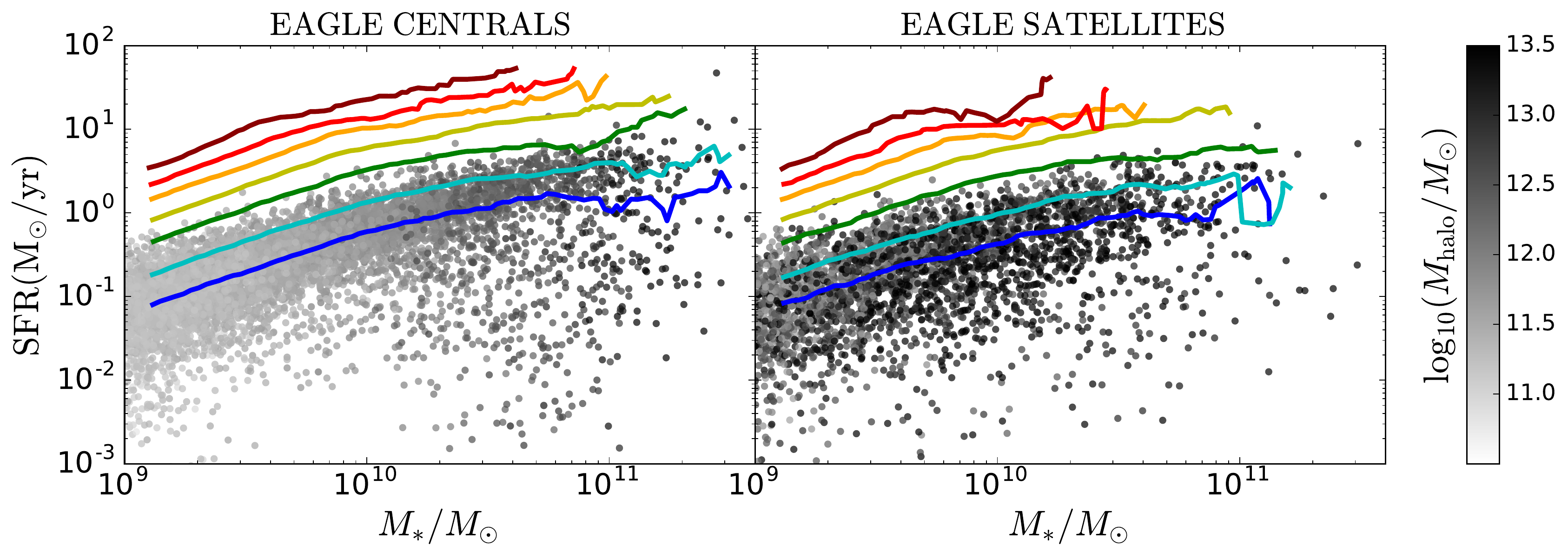}
	\includegraphics[scale=0.45]{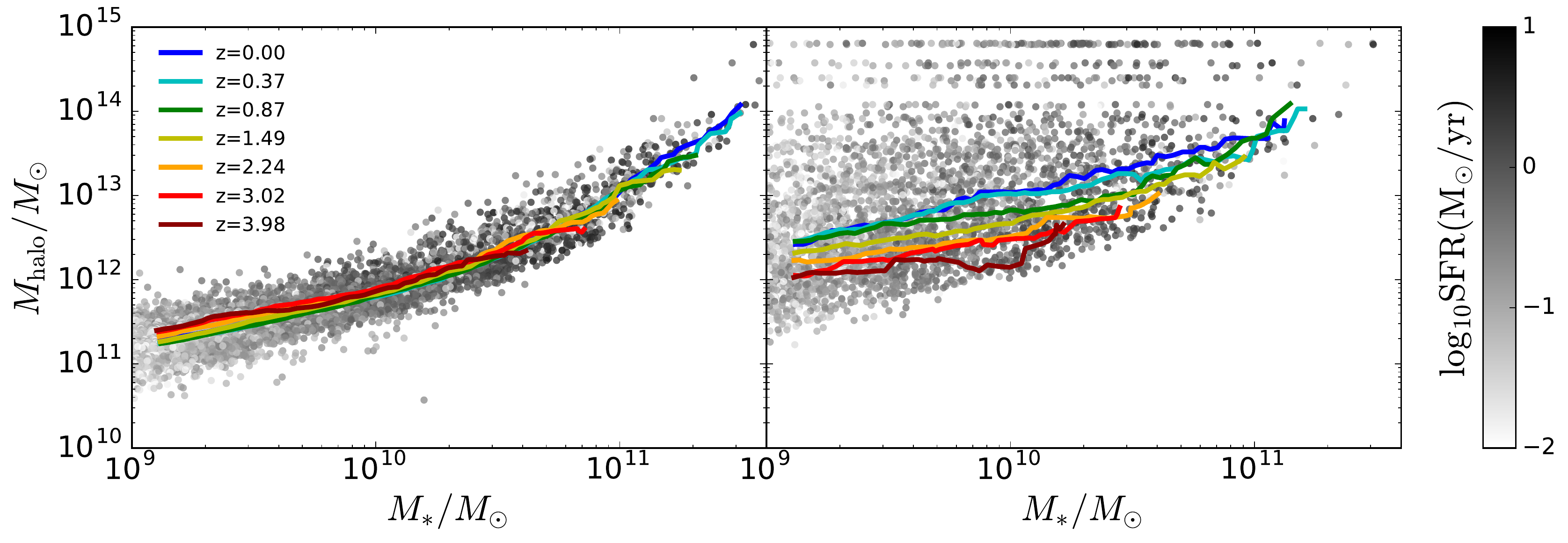}
	\caption{Top: the positions of EAGLE central (left) and satellite (right) galaxies in the stellar mass - star-formation rate plane at $z=0$, colour-coded by halo mass. Bottom: the same galaxies plotted in the stellar mass-halo mass plane, colour-coded by star-formation rate. On all panels, we overplot the evolution in the median relation with redshift, textcolor{red}{using a 0.25dex running median}. At fixed stellar mass, star-formation rates increase substantially towards higher redshift for both central and satellite galaxies. However, the typical halo mass of central galaxies at fixed stellar mass is largely invariant with redshift.}
    \label{fig:EAGLE_dems}
\end{figure*}	
\section{EAGLE galaxies across cosmic time}\label{sec:hm_sm_sfr}
\subsection{Sample selection and galaxy properties}
There are a number of EAGLE simulations available \citep{McAlpine2015}. We draw our galaxy samples from version Ref-L100N1504, due to its large volume (box of side length $100\rm{Mpc}$, comoving) and particle number (7 billion). We select EAGLE galaxies with $M_{*}>10^{9}M_{\odot}$. Large numbers of particles are required to sample the formation history of each galaxy, and EAGLE galaxy properties become unreliable below this stellar mass \citep{McAlpine2015,Schaye2015}. Imposing this stellar mass limit also makes comparison to observational data easier. The minimum $\rm{SFR}$ resolved by EAGLE is $\sim10^{-3}M_{\odot}\rm{yr}^{-1}$ due to gas particle resolution, and some galaxies (<15\% at $z=0$ and fewer at higher redshifts) are assigned $\rm{SFR}=0M_{\odot}\rm{yr}^{-1}$. We exclude these $\rm{SFR}=0M_{\odot}\rm{yr}^{-1}$ galaxies from the PCA analysis described in Section \ref{sec:pca_description}, since we use the logarithm of the SFR (note that our results are largely unchanged if we instead assign these galaxies with a low `limit' star-formation rate). However, we do retain these galaxies in Appendix \ref{sec:hods_from_eagle}, in order to construct the halo occupation distribution of mass-selected sources.\\
\indent We use the total friends-of-friends (FOF) mass of the galaxy's halo \citep{Davis1985}, labelled as GroupMass in the EAGLE FOF table, as opposed to the subhalo mass. We identify central galaxies as those galaxies for which $\rm{SubGroupNumber}=0$, and satellite galaxies as galaxies with $\rm{SubGroupNumber}>0$. The stellar mass and star-formation rates used are those within a 30pkpc (proper, as opposed to comoving, kpc) aperture, taken from the EAGLE Aperture table.
\subsection{Relationships between halo mass, stellar mass and SFR and evolution with redshift}
Observed galaxies have long been found to inhabit a particular region in the stellar mass - star-formation rate plane, often dubbed the `star-forming main sequence' \textcolor{black}{\citep[e.g.][]{Noeske2007,Renzini2015,Lee2015a}}. \textcolor{black}{This broadly linear relation appears to persist with redshift (with evolving normalisation), though its absolute normalisation and slope differ from sample-to-sample \citep[see the compilation of][]{Speagle2014}. The extent to which more passive galaxies occupy a wholly separate region of the plane has also been questioned \citep{Eales2017}}. \\
\indent In Figure \ref{fig:EAGLE_dems} we plot two commonly studied relations as output by EAGLE. In the upper panels, we present the stellar mass vs star-formation rate relation of EAGLE central and satellite galaxies at $z=0$, and overplot the evolution of the median relation back to $z=4$. The evolution of this relation is fairly smooth, with both central and satellite galaxies in the simulations forming stars at a faster rate at higher redshift, for fixed stellar mass. Galaxies at $z=0$ are colour-coded by their group halo mass. For centrals, there is a strong trend that more massive galaxies are hosted by more massive dark matter haloes, as expected. Furthermore, at lower stellar masses ($M_{*}<10^{10}M_{\odot}$) there is a weak trend that (at fixed stellar mass) more highly SF galaxies reside in more massive dark matter haloes; this is discussed in more detail, and found to match observational results, in \cite{Cochrane2017a}. Satellite galaxies inhabit similar regions of this plane, but their halo mass appears to correlate less strongly with position. \\
\indent The lower panels of Figure \ref{fig:EAGLE_dems} show the stellar mass vs halo mass relation. The relationship between stellar mass and halo mass reflects the time-integrated efficiency of stellar mass growth relative to halo growth. As found in many other studies, the host dark matter halo mass to stellar mass relation does not evolve with redshift for central galaxies. This could be because star formation in galaxies tracks the specific mass accretion rate of the halo \citep{Rodriguez-Puebla2016,Cochrane2017}. \\
\indent Some work has already used EAGLE to study these relations in detail. For example, \cite{Matthee2016} found that the scatter in stellar mass at fixed halo mass decreases with increasing halo mass, from $\sim0.25\,\rm{dex}$ at $M_{\rm{halo}} = 10^{11}M_{\odot}$ to  $\sim0.12\,\rm{dex}$ at $M_{\rm{halo}} = 10^{13}M_{\odot}$, stressing that this scatter is not, as is often assumed, independent of halo mass. They attributed some of this scatter (up to $\sim0.04\,\rm{dex}$) to the halo formation time, but found no dark matter halo property that can account for the remaining scatter. In this paper, we look at the role of star-formation rate in driving this scatter.\\
\indent The relation between halo mass and the galactic content of a halo (in terms of the stellar mass and SFR distributions of the constituent galaxies) is often described by HOD modelling. HOD models parametrize the number of central and satellite galaxies in a halo as a function of halo mass. The results of HOD modelling can be strongly dependent on the form of the parametrization adopted \citep[e.g. see][]{Contreras2013}; current parametrizations have primarily been devised for studies of galaxy populations above some stellar mass or star-formation rate limit, and it is not clear whether these are appropriate for other galaxy samples, such as those selected within stellar mass or star-formation rate bins. In Appendix A, we use the EAGLE samples developed here to investigate this. We find that the flexible parametrization proposed by \cite{Geach2012} and used by \cite{Cochrane2017,Cochrane2017a} is able to provide a good description of the halo occupancy for a wide range of galaxy selection criteria.

\begin{table*}
\begin{center}
\begin{tabular}{c|c|c|c|c|c|c}
{\bf{Halo mass range}} & {\bf{PC1}} & {\bf{Var1}} & {\bf{PC2}} & {\bf{Var2}} & {\bf{PC3}} & {\bf{Var3}} \\

\hline
\multicolumn{6}{l}{{Whole EAGLE sample, $M_{*}>10^{9}M_{\odot}$, $\rm{SFR}>0M_{\odot}\rm{yr}^{-1}$}} \\
\hline
$10^{10}M_{\odot}<M_{\rm{halo}}<10^{14}M_{\odot}$ & [0.52, 0.65, 0.55] & 63.2\% & [0.75, -0.04, -0.66] & 25.8\% & [0.41, -0.76, 0.51] & 11.0\% \\
\hline
\multicolumn{6}{l}{{EAGLE central galaxies}} \\
\hline
$10^{10}M_{\odot}<M_{\rm{halo}}<10^{12}M_{\odot}$, \textcolor{black}{$M_{*}>10^{9}M_{\odot}$} & [0.59, 0.60, 0.54] & 78.6\% & [0.46, 0.31, -0.83] & 14.4\% & [0.67, -0.74, 0.09] & 7.0\% \\
$10^{12}M_{\odot}<M_{\rm{halo}}<10^{14}M_{\odot}$, \textcolor{black}{$M_{*}>10^{9}M_{\odot}$} & [0.71, 0.71, 0.04] & 61.1\% & [0.07, -0.01, -1.0] & 33.4\% & [0.71, -0.71, 0.06] & 5.5\% \\
\textcolor{mygray}{$10^{10}M_{\odot}<M_{\rm{halo}}<10^{12}M_{\odot}$, $M_{*}>10^{10}M_{\odot}$} & \textcolor{mygray}{[0.67, 0.62, 0.41]} & \textcolor{mygray}{47.7\%} & \textcolor{mygray}{[0.15, 0.43, -0.89]} & \textcolor{mygray}{30.8\%} & \textcolor{mygray}{[0.73, -0.65, -0.20]} & \textcolor{mygray}{21.5\%}\\
\textcolor{mygray}{$10^{12}M_{\odot}<M_{\rm{halo}}<10^{14}M_{\odot}$, $M_{*}>10^{10}M_{\odot}$} & \textcolor{mygray}{[0.71, 0.71, -0.04]} & \textcolor{mygray}{58.9\%} & \textcolor{mygray}{[0.04, -0.10, -0.99]} & \textcolor{mygray}{33.6\%} & \textcolor{mygray}{[0.70, -0.70, 0.10]} & \textcolor{mygray}{7.5\%}\\
\hline
\multicolumn{6}{l}{{EAGLE satellite galaxies}} \\
\hline
$10^{10}M_{\odot}<M_{\rm{halo}}<10^{12}M_{\odot}$, \textcolor{black}{$M_{*}>10^{9}M_{\odot}$} & [0.53, 0.61, 0.59] &  58.1\% & [0.84, -0.28, -0.46] & 23.7\% & [0.11, -0.74, 0.66] & 18.2\% \\
$10^{12}M_{\odot}<M_{\rm{halo}}<10^{14}M_{\odot}$, \textcolor{black}{$M_{*}>10^{9}M_{\odot}$} & [0.39, 0.70, 0.60] & 54.9\% & [0.84, -0.02, -0.54] & 32.6\% & [0.37, -0.71, 0.60] & 12.5\% \\
\textcolor{mygray}{$10^{12}M_{\odot}<M_{\rm{halo}}<10^{14}M_{\odot}$, $M_{*}>10^{10}M_{\odot}$} & \textcolor{mygray}{[0.72, 0.68, -0.09]} & \textcolor{mygray}{43.5\%} & \textcolor{mygray}{[0.24, -0.37, -0.90]} & \textcolor{mygray}{36.4\%} & \textcolor{mygray}{[0.65, -0.63, 0.44]} & \textcolor{mygray}{20.2\%} \\
\hline
\multicolumn{6}{l}{{SDSS central galaxies}} \\
\hline
$10^{12}M_{\odot}<M_{\rm{halo}}<10^{14}M_{\odot}$, \textcolor{black}{$M_{*}>10^{9}M_{\odot}$} & [0.71, 0.71, 0.03] & 55.5\% & [0.09, -0.05, -0.99] & 33.5\% & [0.70, -0.70, 0.10] & 11.0\% \\
\textcolor{mygray}{$10^{12}M_{\odot}<M_{\rm{halo}}<10^{14}M_{\odot}$, $M_{*}>10^{10}M_{\odot}$} & \textcolor{mygray}{[0.71, 0.71, 0.03]} & \textcolor{mygray}{55.5\%} & \textcolor{mygray}{[0.09, -0.05, -0.99]} & \textcolor{mygray}{33.5\%} & \textcolor{mygray}{[0.70, -0.70, 0.10]} & \textcolor{mygray}{11.0\%} \\
\hline
\multicolumn{6}{l}{{SDSS satellite galaxies}} \\
\hline
$10^{12}M_{\odot}<M_{\rm{halo}}<10^{14}M_{\odot}$, \textcolor{black}{$M_{*}>10^{9}M_{\odot}$} & [0.48, 0.69, 0.54] & 51.4\% & [0.76, -0.02, -0.65] & 31.7\% & [-0.44, 0.72, -0.53] & 16.9\% \\
\textcolor{mygray}{$10^{12}M_{\odot}<M_{\rm{halo}}<10^{14}M_{\odot}$, $M_{*}>10^{10}M_{\odot}$} & \textcolor{mygray}{[0.60, 0.71, 0.37]} & \textcolor{mygray}{47.4\%} & \textcolor{mygray}{[0.52, 0.00, -0.85]} & \textcolor{mygray}{33.3\%} & \textcolor{mygray}{[-0.60, 0.71, -0.37]} & \textcolor{mygray}{19.3\%} \\
\hline
\end{tabular}
\caption{Principal components of EAGLE galaxies at z=0.00 (snapshot 28), with vectors signifying $[\log_{10}M_{\rm{halo}}/M_{\odot},\,\, \log_{10}M_{*}/M_{\odot},\,\, \log_{10}\rm{SFR}/M_{\odot}\rm{yr}^{-1}]$. Central and satellite galaxies with $M_{*}>10^{9}M_{\odot}$ and $\rm{SFR}>0M_{\odot}\rm{yr}^{-1}$ were included in the analysis, though there are no significant differences in the principal components when less massive galaxies are included, or if $\rm{SFR}=0\,M_{\odot}\rm{yr}^{-1}$ galaxies are included at some low SFR limit. Rows highlighted in grey represent principal components calculated using solely high mass galaxies ($M_{*}>10^{10}M_{\odot}$; see Section \ref{sec:two_sm}). There is little cosmic evolution in the variance contained by each component (see Figure \ref{fig:EAGLE_pca}), so we detail only this one redshift, presenting others in Appendix \ref{sec:appendix1}. We also present the principal components for SDSS galaxies in high mass haloes, which show strong agreement with the simulated galaxies.}
\label{table:pca_table_whole}
\end{center}
\end{table*}
\section{Distinguishing the roles of $M_{\rm{halo}}$, $M_{*}$ and SFR using Principal Component Analysis}\label{sec:pca_overall}
Principal Component Analysis (PCA) is a statistical approach used to describe the variance within a dataset. Observed variables - here, halo mass, stellar mass and star-formation rate - are converted into a set of uncorrelated variables, the orthogonal principal components. The first component reveals the direction of maximum variance. Successive components contain less of the variance of the population. This way, some latter components may be dominated by noise, leaving the data decomposed into fewer dimensions. \\
\indent PCA has been used in a number of recent galaxy evolution studies. \cite{Bothwell2016} selected (mostly low redshift) galaxies with cold gas measurements, arguing that the relation between stellar mass, molecular gas mass and gas-phase metallicity is more fundamental than the traditional `Fundamental Metallicity Relation' \citep{Mannucci2010} which uses star-formation rate rather than molecular gas mass. \cite{Lagos2015} used PCA to show that EAGLE galaxies occupy a nearly flat surface within the neutral gas - stellar mass - star-formation rate plane, with little redshift evolution. Neither of these studies look at the role of halo mass, nor is environment studied in great detail in the follow-up work of \cite{Hashimoto2018}.\\
\indent In the following subsections, we identify the principal components within the 3 parameters of halo mass, stellar mass, and star-formation rate, for central and satellite galaxies within the EAGLE simulation. We also investigate the differences between the principal components of galaxies hosted by low mass haloes ($10^{10}-10^{12}M_{\odot}$) and high mass haloes ($10^{12}-10^{14}M_{\odot}$). This roughly splits haloes into those above and below the peak of the stellar mass - halo mass relation (SHMR), which quantifies the efficiency of stellar mass build-up as a function of dark matter halo mass \citep[e.g.][]{Moster2013,Behroozi2013}. 
\subsection{PCA procedure}\label{sec:pca_description}
\indent Principal Component Analysis describes data in terms of linear combinations of the input variables. Therefore, we take the logarithm of all three quantities, supplying vectors of the form $[\log_{10}M_{\rm{halo}}/M_{\odot},\,\, \log_{10}M_{*}/M_{\odot},\,\, \log_{10}\rm{SFR}/M_{\odot}\rm{yr}^{-1}]$. We use the PCA python tool {\it{scikit.learn}} to perform the PCA analysis. Each variable is normalised to its mean \textcolor{black}{and scaled to unit variance} for each galaxy sample input to the PCA.

\begin{figure*} 
	\centering
	\includegraphics[scale=0.62]{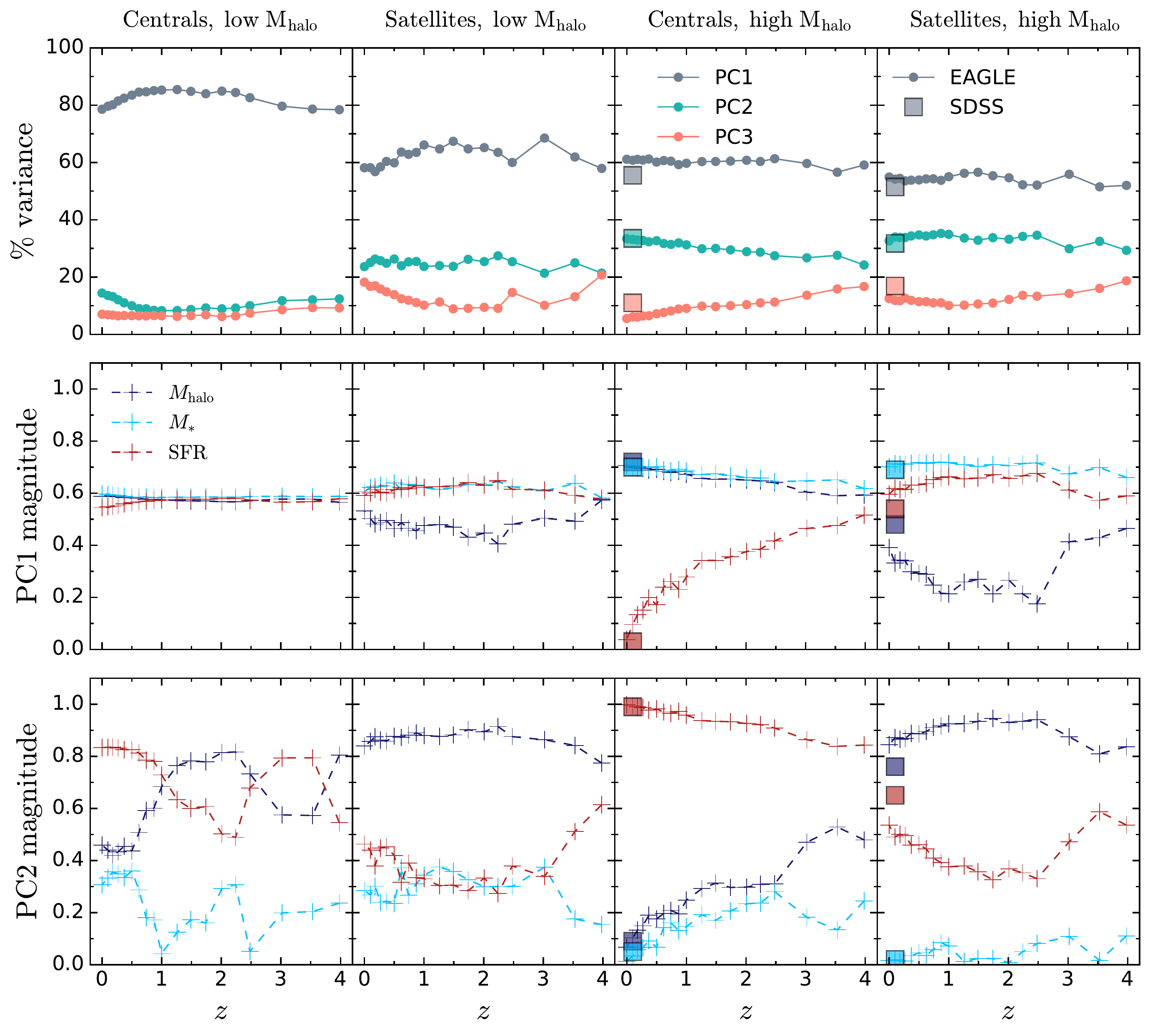}
	\caption{Top: Evolution in the variance contained by the three principal components of EAGLE galaxies, split into centrals and satellites, and into high and low mass haloes. There is remarkably little evolution back to $z=4$ for either central or satellite galaxies. Middle and bottom: The magnitudes of the vectors of the first two principal components for EAGLE galaxies at each redshift, split as above. Again, there is little evolution in these, apart from the decoupling of star-formation rate from the stellar mass and halo mass for central galaxies in high mass haloes at low redshifts (and PC2 for centrals in low mass haloes, which is noisy due to low variance in this component). The square symbols show data points for SDSS galaxies in high mass haloes. These are in very good agreement with the EAGLE results at $z\sim0$.}
    \label{fig:EAGLE_pca}
\end{figure*}


\subsection{The whole EAGLE sample at $z=0$}\label{sec:whole_and_sub}
Initially, we perform PCA on our whole sample of EAGLE galaxies \textcolor{black}{with $M_{*}>10^{9}M_{\odot}$}, within the halo mass range $M_{\rm{halo}}=10^{10}-10^{14}M_{\odot}$ at $z=0$. \textcolor{black}{Our two halo mass bins are $10^{10}M_{\odot}<M_{\rm{halo}}<10^{12}M_{\odot}$ and $10^{12}M_{\odot}<M_{\rm{halo}}<10^{14}M_{\odot}$, but note that, because of the stellar mass cut applied to select only well-resolved galaxies, most of our haloes in the mass range $10^{10}M_{\odot}<M_{\rm{halo}}<10^{12}M_{\odot}$ are actually at $M_{\rm{halo}}>10^{11}M_{\odot}$.} The resulting principal components are provided in Table \ref{table:pca_table_whole}. The primary relation is a positive correlation between halo mass, stellar mass and star-formation rate. This axis encapsulates the majority ($\sim63\%$) of the sample variance. The secondary component is a negative correlation between halo mass and star-formation rate, with little dependence on stellar mass. This reflects the tendency of galaxies in high mass haloes to have low star-formation rates, broadly independent of their stellar mass, and is suggestive of environmental quenching. \\
\indent Next, we divide the galaxies into four subsamples, splitting by central/satellite galaxy and by halo mass \textcolor{black}{but retaining the $M_{*}>10^{9}M_{\odot}$ stellar mass cut. \footnote{We have tested the impact of this stellar mass cut, and find that including galaxies with lower stellar masses (e.g. imposing a lower limit of $M_{*}=10^{8}M_{\odot}$), where host halo masses are typically lower, makes little difference to our results.}} We find that the principal components vary between the four subsamples (see Table \ref{table:pca_table_whole} for full details of the $z=0$ principal components). We summarise the results here.\\
\begin{itemize}
\item[-]For $z=0$ central galaxies in low mass haloes \textcolor{black}{($10^{10}M_{\odot}<M_{\rm{halo}}<10^{12}M_{\odot}$)}, $\sim79\%$ of the variance of the population is contained in PC1, which represents a positive correlation between halo mass, stellar mass and star-formation rate. Note that the star-formation rate is a key component in this, i.e. we don't just obtain a halo mass - stellar mass component, nor do we obtain two separate components that encode the halo mass - stellar mass and the stellar mass - SFR correlations. PC2, which contains a comparatively small $\sim14\%$ of the variance, reflects the secondary negative correlation between star-formation rate and the other two parameters. This is significantly smaller for the centrals in low mass haloes than for the $z=0$ EAGLE sample as a whole, reflecting the low passive galaxy fraction of this subsample. \\
\item[-]For $z=0$ central galaxies in high mass haloes \textcolor{black}{($10^{12}M_{\odot}<M_{\rm{halo}}<10^{14}M_{\odot}$)}, the primary relation is solely between halo mass and stellar mass ($\sim61\%$), with essentially no component of SFR. PC2 then represents SFR only, containing $33\%$ of the scatter. In high mass haloes, the SFR of the central galaxy thus appears to be decoupled from its stellar mass and halo mass. Note here that since the SFR correlates with neither stellar mass nor halo mass, it is not possible to tell from this alone whether the quenching of star formation for centrals in high mass haloes is driven by stellar or halo mass. We return to this question in Section \ref{sec:two_sm}. \\
\item[-]The first principal component of satellite galaxies in low mass haloes \textcolor{black}{($10^{10}M_{\odot}<M_{\rm{halo}}<10^{12}M_{\odot}$)} is again between halo mass, stellar mass and star-formation rate, though less variance is contained in this component than for the central galaxies in haloes of the same mass ($\sim58\%$ compared to  $\sim79\%$). This is likely to be due to the smaller role of the group halo compared to the subhalo in the growth of the satellite galaxy. Indeed, if the subhalo mass is used instead of halo mass in the analysis, then principal components similar to those of the central galaxies are recovered. PC2 indicates scatter in the halo mass - star-formation rate relation ($\sim24\%$), and PC3 is the scatter in the stellar mass - halo mass relation ($\sim18\%$). \\
\item[-]For satellite galaxies in high mass haloes \textcolor{black}{($10^{12}M_{\odot}<M_{\rm{halo}}<10^{14}M_{\odot}$)}, the primary correlation is between stellar mass and star-formation rate (55\%). Although halo mass is also positively correlated with these two, it has a much weaker contribution, probably reflecting the history of the satellites, which formed most of their mass prior to accretion onto a more massive dark matter halo. PC2 ($33\%$) is driven by the negative correlation between halo mass and star-formation rate. Stellar mass does not contribute to this component. This clearly reflects the important role of halo environment, rather than stellar mass, in quenching star formation in satellite galaxies.\\\\
\indent \textcolor{black}{We have tested changing the halo mass threshold between high and low halo mass samples. The change in principal components is quite gradual with halo mass, and our results are insensitive to the exact threshold selected.}
\end{itemize}
\subsection{Comparison to SDSS $z\sim0$ galaxies}\label{sec:sdss_comparison}
To compare our results from the EAGLE simulation with observations, we select galaxies with $M_{*}>10^{9}M_{\odot}$ from the 7th data release (DR7; \citealt{Abazajian2009}) of the Sloan Digital Sky Survey (SDSS; \citealt{York2000}). We draw stellar masses and star-formation rates from the value-added spectroscopic catalogues produced by MPA-JHU\footnote{http://wwwmpa.mpa-garching.mpg.de/SDSS/DR7/} \citep{Kauffmann2003,Brinchmann2004}. We obtain halo mass and central/satellite estimates from the group catalogues of \cite{Yang2007}. These primarily ascribe halo masses of $M_{\rm{halo}}>10^{12}M_{\odot}$, so we can only reliably compare these observational data with simulated EAGLE galaxies in high mass haloes. Our final sample consists of 319,158 SDSS galaxies at $z<0.2$.\\
\indent The populations of EAGLE and SDSS galaxies are not perfectly matched, with EAGLE galaxies having lower masses and star-formation rates, on average, than the observed SDSS galaxies. This is in part because the lowest mass (hence, broadly, lowest luminosity) galaxies in SDSS will only be detectable at the lowest redshifts, and hence over a smaller observed volume than is available to higher mass (luminosity) galaxies. It is also well-known that the specific star-formation rates of EAGLE star-forming galaxies are $0.2-0.5\,\rm{dex}$ below those inferred from observations, across all redshifts \citep{Furlong2015a}. Nevertheless, despite these small inconsistencies in the distributions and absolute values of stellar mass and star-formation rate, we are still able to make comparisons between the simulations and our data. This is because the PCA approach considers the broad trends between stellar mass, star-formation rate and halo mass, and it is therefore not necessary to select a sample of galaxies from EAGLE that matches the observed population exactly. \textcolor{black}{For the same reasons, we find that applying different redshift cuts to the SDSS sample, to generate a sample better matched in stellar mass distribution, does not change the principal components significantly. Thus, given that this would only reduce the sample size, we choose not to apply a further redshift selection to the SDSS data.} \\
\indent We perform exactly the same analysis for SDSS galaxies as for EAGLE and find excellent agreement between the principal components of the observational and simulated data for both satellites and centrals in high mass haloes at $z\sim0$ (see Table \ref{table:pca_table_whole} and Figure \ref{fig:EAGLE_pca}). For observed central galaxies in high mass haloes, the first principal component embodies the positive correlation between halo mass and stellar mass, with star-formation rate decoupled from this as the second principal component. For observed satellite galaxies in high mass haloes, the key relation is between all three variables, but the secondary component, which contains $\sim32\%$ of the variance, is the negative correlation between halo mass and star-formation rate. Both the components and the magnitudes of the variance they contain are very similar to those found in EAGLE, given the same stellar mass, halo mass and central/satellite sample selections. Thus, we are confident in the conclusions that we draw from EAGLE. This strong agreement between SDSS and EAGLE also gives us further confidence in the viability of the EAGLE HOD modelling in Appendix \ref{sec:hods_from_eagle}.

\begin{figure*} 
	\centering
	\includegraphics[scale=0.9]{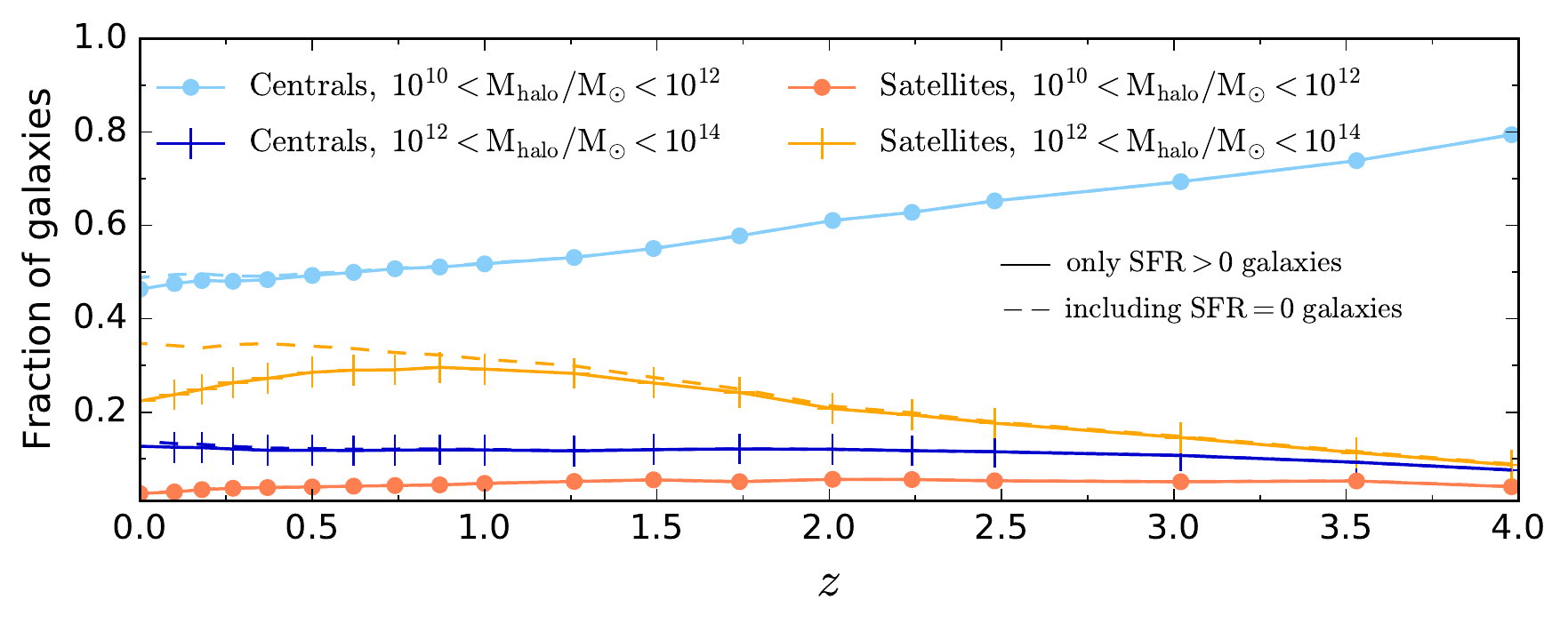}
	\includegraphics[scale=0.9]{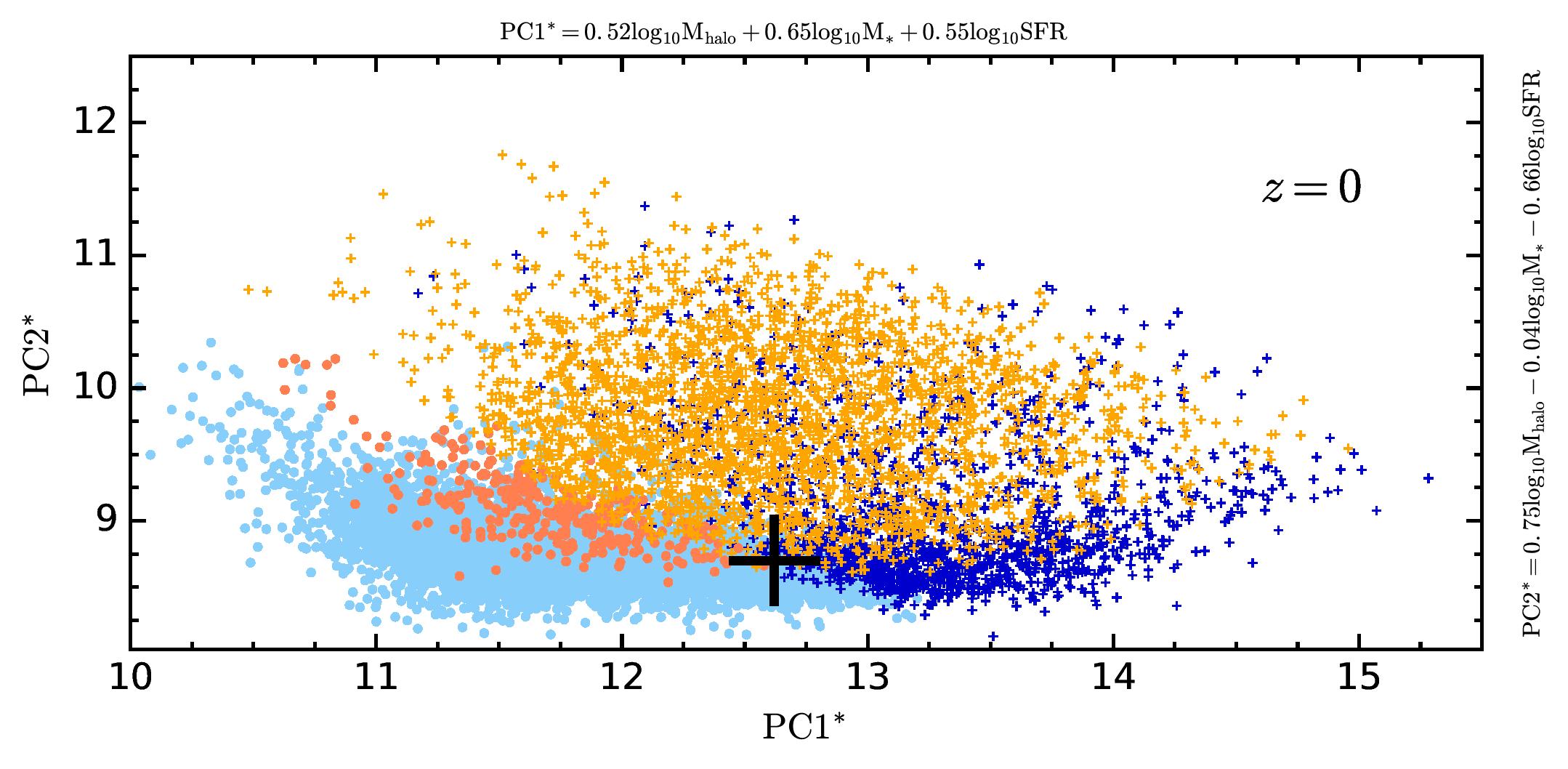}
	\includegraphics[scale=0.93]{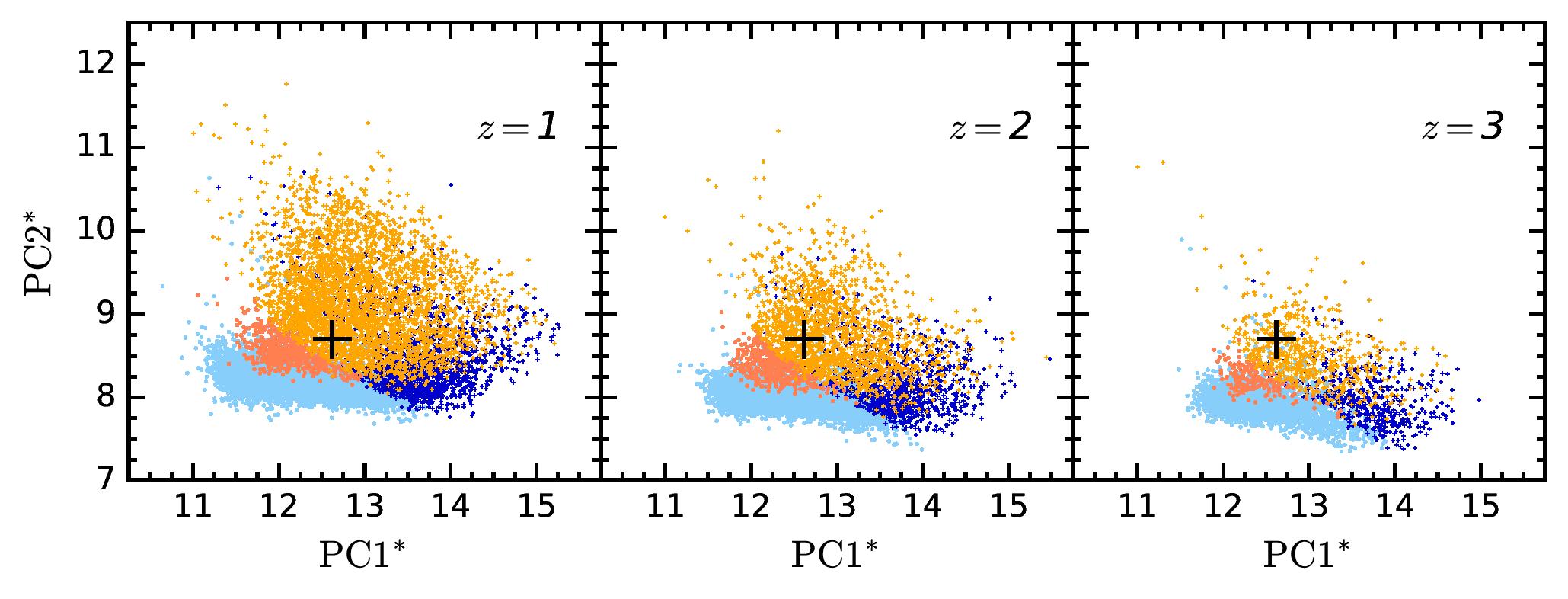}
	\caption{Top: the cosmic evolution of the distribution of $M_{*}>10^{9}M_{\odot}$ galaxies between centrals and satellites and between low ($10^{10}<M_{\rm{halo}}/M_{\odot}<10^{12}$) and high ($10^{12}<M_{\rm{halo}}/M_{\odot}<10^{14}$) mass haloes. Solid lines represent the $\rm{SFR} > 0M_{\odot}\rm{yr}^{-1}$ population. The vast majority of EAGLE galaxies fall into this category. Dashed lines represent the whole EAGLE population, including $\rm{SFR}=0M_{\odot}\rm{yr}^{-1}$ galaxies. The only population with significant numbers of $\rm{SFR}=0M_{\odot}\rm{yr}^{-1}$ galaxies is the satellite galaxies in high mass haloes; these grow in number significantly below $z\sim1$. Middle: All four $z=0$ EAGLE samples are plotted on the axes of the first two principal components of the whole $z=0$ EAGLE sample ($M_{*}>10^{9}M_{\odot}$, $10^{10}<M_{\rm{halo}}/M_{\odot}<10^{14}$). \textcolor{black}{The black cross shows the approximate meeting point of the populations at $z=0$, to guide the eye.} Bottom: The same plot for EAGLE galaxies at different redshifts. The numbers of galaxies in each of the four samples changes, but the typical positions of the four subsamples on the PCA plot do not (apart from moving upwards towards lower redshifts as typical star-formation rates decrease, \textcolor{black}{as shown by the relative position of the $z=0$ black cross}). We find that the principal correlations between $M_{\rm{halo}}$, $M_{*}$ and $\rm{SFR}$ are fundamental, and independent of cosmic time.}
    \label{fig:pca_project}
\end{figure*}	
\begin{figure*} 
	\centering
	\includegraphics[scale=0.61]{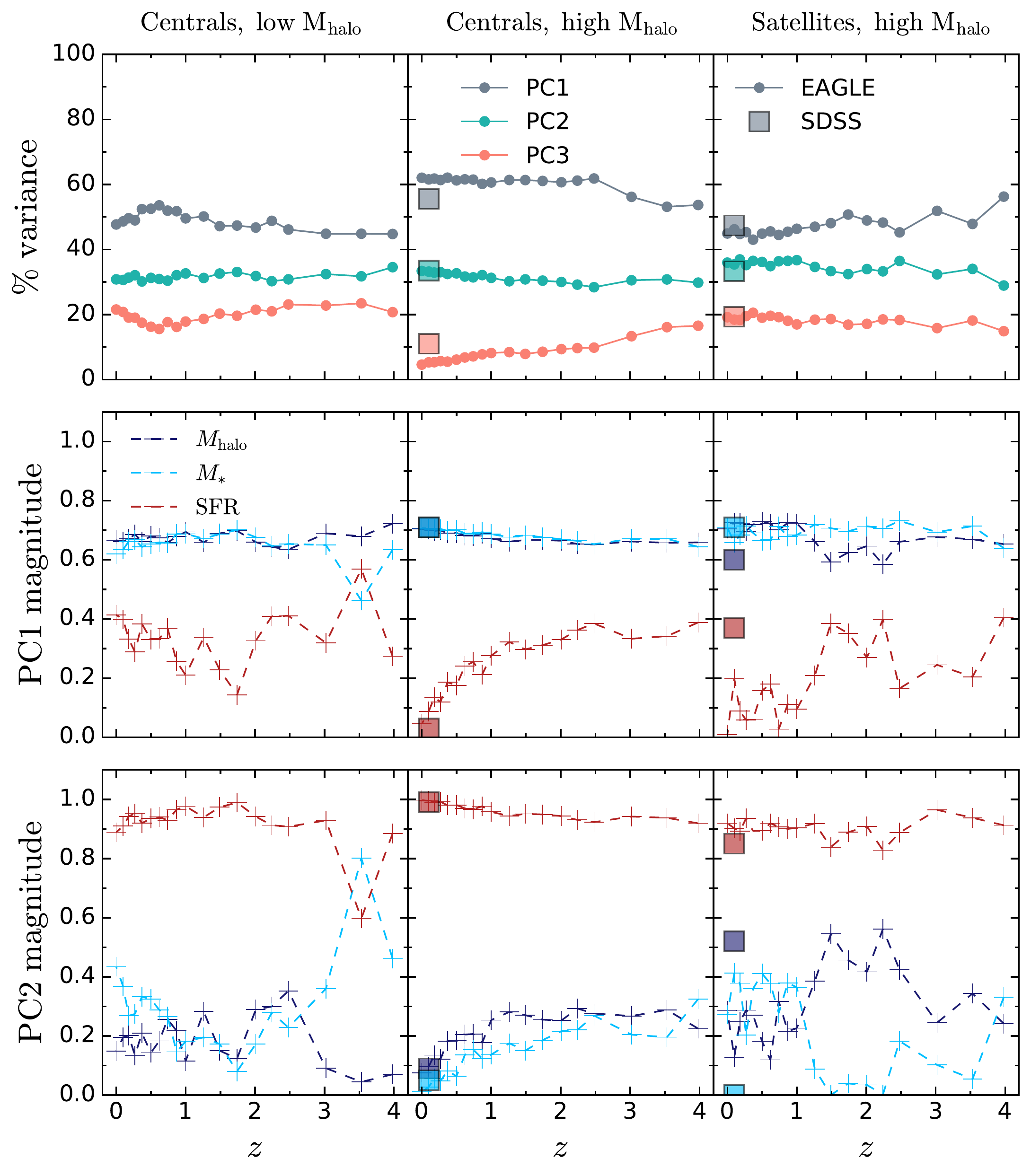}
	\caption{Evolution in the variance contained by the three principal components of high mass EAGLE galaxies ($M_{*}>10^{10}M_{\odot}$) in different environments. For these high stellar mass subsamples, the principal components of the central galaxies in high and low mass haloes and of the satellite galaxies in high mass haloes are very similar (there are insufficient high mass satellites in low mass haloes to investigate these). Star-formation rate is largely decoupled from the stellar mass - halo mass relation, apparently due to processes related to stellar mass but not halo environment. There is little redshift evolution for the samples in any of the halo environments. Again, the SDSS data points agree fairly well with EAGLE.}
    \label{fig:EAGLE_pca_high_sm}
\end{figure*}

\begin{figure} 
	\centering
	\includegraphics[scale=0.55]{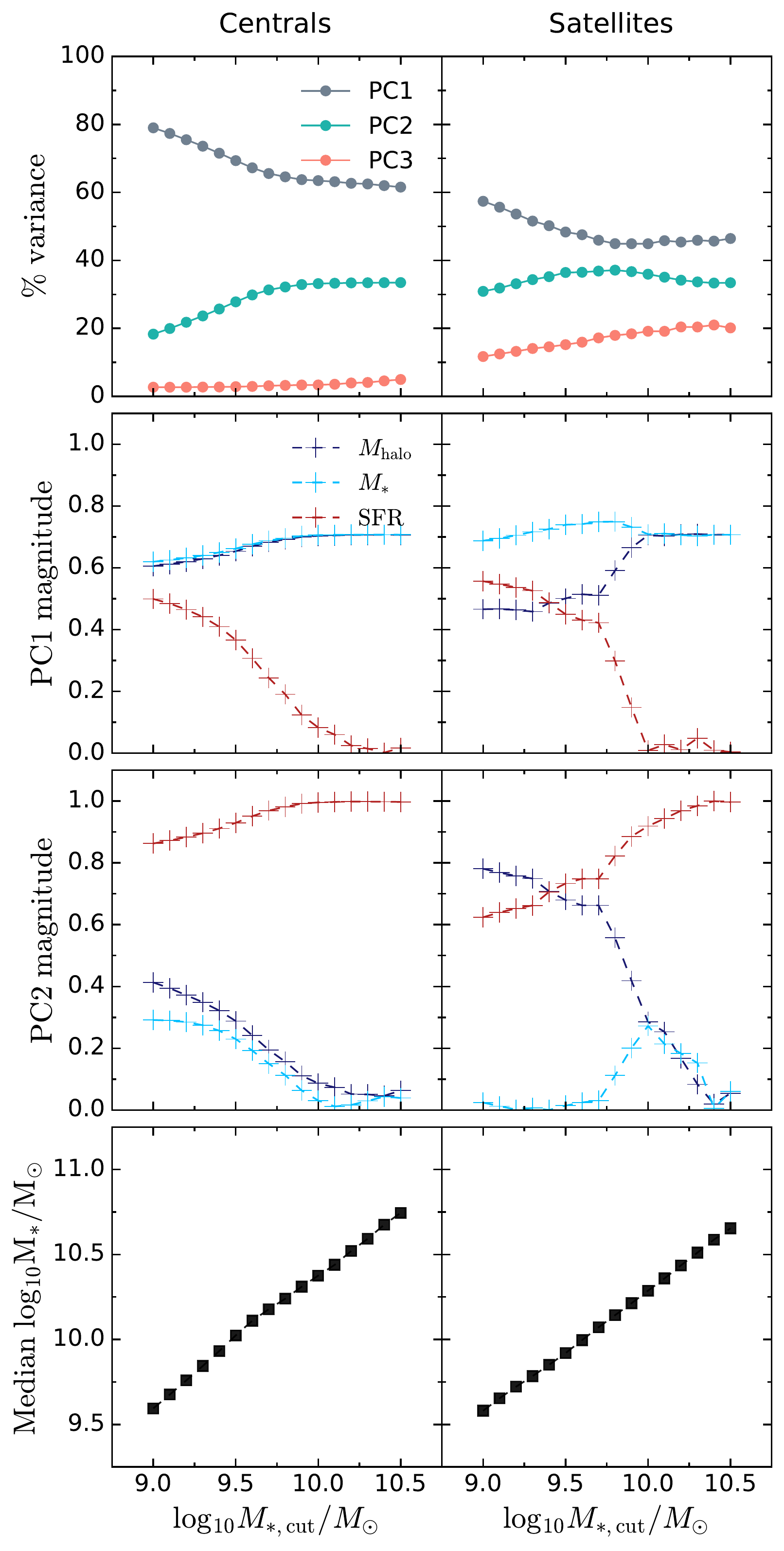}
	\caption{\textcolor{black}{Principal components of subsamples of central (left) and satellite (right) galaxies as a function of minimum stellar mass. EAGLE galaxies at $z=0$ with $M_{*}>M_{*,\rm{cut}}$ in the halo mass range $10^{10}M_{\odot}<M_{\rm{halo}}<10^{14}M_{\odot}$ are included in each bin. The median stellar mass of each subsample is plotted in the bottom panel. Star-formation rate decouples from stellar mass and halo mass above $M_{*}\sim10^{10}M_{\odot}$.}}
    \label{fig:sm_threshold}
\end{figure}

\subsection{PCA evolution with redshift}\label{sec:pca_z_ev}
EAGLE catalogues span a wide range in redshifts. Therefore, it is possible to study the evolution of the principal components over cosmic time. We repeat the principal component analysis at all EAGLE redshifts back to $z\sim4$ (Appendix \ref{sec:appendix1} provides a full table of results). It is remarkable how consistent both the principal components and the variances are for most of the samples. We show the evolution in the variance contained by each principal component in the top panel of Figure \ref{fig:EAGLE_pca}. There is \textcolor{black}{little evolution in these values}, at fixed halo mass within the central and satellite galaxy population. In the middle and bottom panels, we plot the magnitudes of each component of the vectors themselves for the first and second principal components. These, too, show little evolution in most cases. One exception is PC2 of centrals in low-mass haloes, but this is simply noisy due to low variance in that principal component. A second exception is the SFR component of PC1 for central galaxies in high mass haloes. In the higher redshift slices of EAGLE, the star-formation rate of central galaxies in high mass haloes is positively correlated with their stellar mass and host halo mass very similarly to that of lower-mass haloes. However, the star formation becomes increasingly decoupled from the halo and stellar mass towards low redshift. Interestingly, this seems not to occur for central galaxies in lower mass haloes; the positive $M_{\rm{halo}}$, $M_{*}$, $\rm{SFR}$ relation of PC1 holds to $z=0$ with little change in the magnitudes of the principal components, and there is only a small decrease ($<8\%$) in the percentage of variance contained by PC1 since $z=1$. \\
\indent The lack of evolution in the PCA view of satellite galaxies is also interesting, given that the percentage of passive galaxies evolves so strongly at low redshift, particularly at low stellar masses \citep[see, for example, the stellar mass functions of][]{Moutard2016}. Our results indicate that the mechanism of environment quenching does not evolve with redshift. This is in line with \cite{Peng2010}, who find that the environmental quenching efficiency as a function of overdensity is invariant with redshift back to $z=1$. \\
\indent Figure \ref{fig:pca_project} presents a complementary view of the evolution of these different populations of galaxies. In the top panel, we show the fraction of the total sample that are central and satellite galaxies in haloes of different masses, as a function of redshift. The fraction of galaxies that are satellites in high mass haloes increases significantly, from $<10\%$ at $z\sim4$ to $\sim30\%$ at $z\sim0$. From $z\sim4$ to $z\sim1$, this reflects increasing numbers of star-forming satellites. Below $z\sim1$, there are a large number of $\rm{SFR}=0M_{\odot}yr^{-1}$ satellite galaxies in massive haloes (around a third of EAGLE satellite galaxies in high mass haloes have unresolved SFRs). \\
\indent In the lower panels of Figure \ref{fig:pca_project} we plot our EAGLE subsamples on the PC1-PC2 plane defined by the whole sample at $z=0$, as given in Table \ref{table:pca_table_whole}. The middle panel shows the different regions of the plane that these populations occupy at $z=0$. Each subsample occupies a fairly well-defined region of the plane. We do not show EAGLE galaxies with low star-formation rates that are unresolved by EAGLE and assigned $\rm{SFR}=0M_{\odot}yr^{-1}$, since their exact SFRs are unknown. Depending on the exact SFR adopted, these will lie towards the upper-left corner of the PC1-PC2 plot, naturally extending the plotted distribution of high-mass halo satellites. \\
\indent In the lower panels, we show examples of the same plot at different redshifts from EAGLE, \textcolor{black}{with the rough meeting point of the four populations at $z=0$ shown by a black cross.} Although the numbers of galaxies within the different classes change significantly, there is little redshift evolution in the regions of the plane occupied by galaxies within the same class, save for an overall shift upwards and to the left towards lower redshifts. This reflects typical star-formation rates dropping with cosmic time. 

\subsection{Evidence for stellar mass quenching?}\label{sec:two_sm}
\cite{Peng2010} argues that mass quenching dominates the quenching of massive galaxies at $M_{*}>10^{10.2}M_{\odot}$ (with the stellar mass threshold decreasing slightly towards higher redshift). If, at these high stellar masses, the role of environment is minimal, we might expect the principal components of very massive galaxies to be different. However, any such trend will be hidden in the analysis of Section \ref{sec:whole_and_sub}, because the most massive galaxies are greatly outweighed by the lower mass galaxies which dominate the stellar mass function (except for central galaxies in high mass haloes, which are mostly high mass due to the strong $M_{\rm halo}-M_{*}$ correlation). Therefore, to probe the role of stellar mass in more detail, we select a `high stellar mass' subsample of EAGLE galaxies with $M_{*}>10^{10}M_{\odot}$, and repeat the analysis on this subsample.\\
\indent We present the principal components of high mass EAGLE galaxies in Figure \ref{fig:EAGLE_pca_high_sm}. Note that we do not show high mass satellite galaxies in low mass haloes, due to their scarcity. It is clear that the principal components of the central galaxies in high and low mass haloes and of the satellite galaxies in high mass haloes are extremely similar, once this stellar mass cut is made. For all three subsamples, PC1 is dominated by the correlation between halo mass and stellar mass. While star-formation rate makes a fairly small contribution towards PC1, it completely dominates PC2, reflecting the decoupling of the star-formation rate from the coevolving stellar mass and halo mass. This trend is seen across halo environments (indeed, although very noisy, high mass centrals in low mass haloes are also consistent with these principal components), and across cosmic time. Thus, star-formation rate decoupling in high stellar mass galaxies appears to be driven by stellar mass rather than halo environment. \\
\indent \textcolor{black}{Motivated by other studies of stellar mass quenching, we  initially chose a `high stellar mass' threshold of $M_{*}>10^{10}M_{\odot}$. To investigate where stellar mass quenching becomes important, we repeat the principal component analysis for samples of galaxies at $z=0$ selected using different minimum stellar mass thresholds. We present our results in Figure \ref{fig:sm_threshold}. We find that the principal components begin to change at $M_{\rm{cut}}=10^{9.5}M_{\odot}$, where the median stellar mass of the sample is $\sim10^{10}M_{\odot}$. Above $M_{\rm{cut}}=10^{10}M_{\odot}$, the star-formation rate is fully decoupled from both stellar mass and halo mass. Our results suggest that the switch in principal components occurs at $\sim10^{10}M_{\odot}$, which is consistent with the stellar mass at which a significant change in the quenched galaxy fraction occurs.}\\

\section{Discussion of quenching modes}\label{sec:quenching_modes}
\subsection{Environment quenching of satellite galaxies}\label{env_quench}
Our results clearly indicate that halo environment plays an important role in the evolution of galaxies. For the whole sample of $z=0$ EAGLE galaxies, the principal correlation within the population is between halo mass, stellar mass and star-formation rate: more massive galaxies tend to live in higher mass haloes and have higher star-formation rates. However, we identify the second principal component as a negative correlation between halo mass and star-formation rate, with no stellar mass term. This points towards a predominant quenching mechanism that is driven by the halo environment, independent of stellar mass. \\
\indent We find that this second component contains the most variance for satellites in high mass haloes. The first principal component of satellites in high mass haloes is dominated by the stellar mass - star-formation rate correlation; for these galaxies, halo mass is less strongly coupled than for the population as a whole. This reflects the accretion histories of satellites, which have tightly correlated star-formation rate and stellar mass but have not grown stellar mass along with the group dark matter halo, but rather in their sub-halo. The halo mass dominates PC2, acting in opposition to the star-formation rate, indicating that that environment is the dominant driver of quenching for these galaxies. This is in line with \cite{Wetzel2013}, who argue that the majority of $z=0$, $M_{*}<10^{10}M_{\odot}$ passive galaxies were quenched as satellites, either within their current host halo or via pre-processing in another halo. Satellites in low mass haloes have similar principal components, but with a slightly larger contribution from stellar mass to PC2. The principal components and variance for satellites in both low and high mass haloes are fairly constant with redshift (see Figure \ref{fig:EAGLE_pca}), indicating that this halo-driven quenching acts from early times. \\
\subsection{Quenching mechanisms for high mass galaxies}\label{mass_quench}
\indent Comparing the results from Figure \ref{fig:EAGLE_pca} and \ref{fig:EAGLE_pca_high_sm}, it is clear that the principal components of the whole sample of satellite galaxies in high mass haloes (which is dominated by lower stellar mass satellites) and of the high stellar mass only satellite galaxy sample are very different. Satellite galaxies with high stellar masses have star-formation rates decoupled from their stellar masses and halo masses in PC1. The similarity of the principal components of these high stellar mass satellite galaxies in high mass haloes to those of high mass centrals in the same haloes, and to high mass centrals in lower mass haloes, suggests that it is the stellar mass rather than the halo environment which is important in this decoupling. However, in all cases star-formation rate is equally decoupled from the halo mass, so this remains difficult to prove. \\
\indent Here, we consider whether these results could be consistent with work proposing that halo mass is also the underlying mechanism of stellar mass quenching. For this, it is important to consider the assembly histories of galaxies, since high stellar mass satellite galaxies are likely to have spent time forming stars building stellar mass as centrals within high mass haloes. As argued by \cite{Gabor2014}, quenching could have preceded satellite accretion and been driven by the halo mass of the previous halo. Given that the satellite's stellar mass will be tightly correlated with the mass of the previous halo, rather than that of the new halo, past halo mass quenching then looks like stellar mass quenching. \\
\indent For the high stellar mass satellite galaxies, we investigate this by examining their halo histories. At each EAGLE timestep, we identify the progenitors of the $z=0$ galaxies, via the EAGLE `main branch' (see \citealt{McAlpine2015}, but note that our results are the same if we instead manually select the most massive progenitor at each redshift). We find that $97\%$ of $z=0$ high stellar mass satellites in high mass haloes have primary progenitors that were central galaxies at $z<1.5$. We collect the most recent central primary progenitors and perform the same principal component analysis on these (note that they span a range of redshifts, $0.1<z<1.5$, as different galaxies first become satellites at different times). The principal components we find (PC1, PC2, PC3 = [0.71, 0.71, -0.04], [0.02, 0.04, 1.0], [0.71, -0.70, 0.02], Var1, Var2, Var3 = 58.4\%, 33.3\%, 8.3\%) are very similar to those of central galaxies in high mass haloes; thus, at the time that these galaxies became satellites, their star-formation rate was already decoupled from both stellar and halo mass. Therefore, from this population we are unable to determine whether it is stellar mass or halo mass that drives the quenching of star-formation. \\
\indent More insight may be gained by looking at high stellar mass central galaxies in low mass haloes. Compared to low mass galaxies in equally massive haloes, star-formation rate is less strongly coupled to halo mass and stellar mass in PC1 for these galaxies. As for the the other high mass galaxy subsamples, PC2 is dominated by SFR. Since the halo mass is low, these objects appear to present the most direct evidence for stellar mass-driven quenching. \\
\indent However, it is important to consider how galaxies with such unusually high stellar-to-halo mass ratios formed. \cite{Gabor2014} find a population of red central galaxies living in low mass haloes within their simulations, which comprise former satellite galaxies that were ejected from a more massive halo following halo-driven quenching. If this is the case for the bulk of these high stellar mass centrals in low mass haloes, then this would remove evidence for stellar mass being the driving factor. We therefore search the progenitors of EAGLE high mass galaxies in low mass haloes to determine whether our galaxies assembled this way. We find that only $\sim17\%$ of $z=0$ EAGLE galaxies have a primary progenitor at $z<1.5$ that was a satellite. Excluding these galaxies does not lead to a significant change in the principal component analysis. This suggests that the decoupling of star-formation rate in these galaxies is driven more directly by the high stellar masses of the galaxies than by their halo mass (although we cannot fully exclude that some other halo property, which also gives rise to the unusually high stellar mass to halo mass ratio, is responsible). The similarity of the principal components for all of these three high stellar mass samples then suggests that stellar mass driven quenching is important in all high stellar mass galaxies. \textcolor{black}{Our analysis confirms that this becomes significant above $\sim10^{10}M{\odot}$.}
\section{Conclusions}\label{sec:conclusions}
In this paper we study the halo environments of galaxies in the EAGLE simulations, focusing on how dark matter halo mass relates to two key baryonic galaxy properties: stellar mass and star-formation rate. \textcolor{black}{We apply the statistical technique Principal Component Analysis to EAGLE galaxies, with comparison to observational data from SDSS. We also show that the halo occupation for EAGLE galaxies selected by many different stellar mass and star-formation bins/limits can be fitted using a single 6-parameter functional form. Our main results are presented here.} 
\begin{itemize}
\item[-] We find a clear primary correlation between host halo mass, galaxy stellar mass and star-formation rate. This correlation is particularly dominant for central galaxies in low mass haloes. It demonstrates the important role that dark matter haloes play in fuelling star formation in galaxies.\\
\item[-] We find strong evidence for environment-driven quenching in satellite galaxies via an anticorrelation between halo mass and star-formation rate in the second principal component.\\
\item[-] We present evidence for an alternative mass-driven quenching mechanism at high stellar mass. This appears to be independent of environment \textcolor{black}{and to set in at $\sim10^{10}M_{\odot}$}.\\
\item[-] Crucially, we find excellent agreement between the principal components derived for EAGLE simulated galaxies and observed galaxies drawn from SDSS at $z\sim0$, for all sub-populations studied. This gives confidence in the validity of the results derived from EAGLE.\\
\item[-] Probing EAGLE galaxies back to $z=4$, we find that the principal components of galaxies within each class do not evolve significantly with redshift, despite changes in the numbers of galaxies in each class and an overall shift towards lower star-formation rates at low redshifts. \textcolor{black}{The only exception is centrals in high mass haloes. For these galaxies, SFR becomes somewhat more decoupled towards low redshift. The overall redshift-independence of the principal components} suggests that the physical mechanisms driving the evolution of galaxies do not evolve strongly over cosmic time.
\end{itemize}
Overall, it is clear that host dark matter haloes play a key role in fuelling and quenching star-formation in galaxies at all redshifts. We show that this role differs for central and satellite galaxies in low/high mass dark matter haloes. However, within these sub-classes, the principal relations between halo mass, stellar mass and star-formation rate, hold across cosmic time. 
\section*{Acknowledgements}
RKC and PNB acknowledge funding from STFC via a studentship and grant ST/M0011229/1. \textcolor{black}{We thank the anonymous reviewer for helpful suggestions.}\\
\indent Funding for the SDSS and SDSS-II was provided by the Alfred P. Sloan Foundation, the Participating Institutions, the National Science Foundation, the U.S. Department of Energy, the National Aeronautics and Space Administration, the Japanese Monbukagakusho, the Max Planck Society, and the Higher Education Funding Council for England. The SDSS was managed by the Astrophysical Research Consortium for the Participating Institutions. The SDSS web site is www.sdss.org.\\
\indent We acknowledge the Virgo Consortium for making their simulation data available and Stuart McAlpine for advice on their use. The EAGLE simulations were performed using the DiRAC-2 facility at Durham, managed by the ICC, and the PRACE facility Curie based in France at TGCC, CEA, Bruy\`{e}res-le-Ch\^{a}tel.
%
\bibliographystyle{mnras}
\bibliography{Edinburgh}
\appendix
\section{How do galaxies populate dark matter haloes?}\label{sec:hods_from_eagle}
\begin{table*}
\begin{center}
\begin{tabular}{l|c|c|c|c|c|c}
{\bf{Stellar mass range}} & {\bf{$\log_{10}M_{\rm{min}}$}} & {\bf{$\sigma_{\log M}$}} & {\bf{$\alpha$}} & {\bf{$F_{c}^{A}$}} & {\bf{$F_{c}^{B}$}} & {\bf{$F_{s}$}} \\
\hline
$M_{*}/M_{\odot}>10^{9}$ & $11.32 \pm 0.02$ & $0.26 \pm 0.02$ & $0.97 \pm 0.03$ & $0.85 \pm 0.07$ & $0.5 \pm 0.3$ & $0.047 \pm 0.008$ \\
$M_{*}/M_{\odot}>10^{9.5}$ & $11.59 \pm 0.03$ & $0.22 \pm 0.03$ & $0.98 \pm 0.05$ & $0.84 \pm 0.09$ & $0.5 \pm 0.5$ & $0.06 \pm 0.01$ \\
$M_{*}/M_{\odot}>10^{10}$ & $11.90 \pm 0.03$ & $0.20 \pm 0.05 $ & $1.05 \pm 0.14$ & $0.90 \pm 0.18$ & $0.5 \pm 0.9$ & $0.04 \pm 0.02$ \\
$M_{*}/M_{\odot}>10^{10.5}$ & $12.30 \pm 0.05$ & $0.18 \pm 0.06$ & $0.8 \pm 0.3$ & $0.5 \pm 0.4$ & $0.06 \pm 0.11$ & $0.10\pm 0.11$ \\
\hline
$10^{9}<M_{*}/M_{\odot}<10^{9.5}$ &$11.40 \pm 0.03$ & $0.22 \pm 0.01$ & $0.96 \pm 0.04$ & $0.00 \pm 0.00$ & $0.43 \pm 0.04$ & $0.03 \pm 0.01$ \\
$10^{9.5}<M_{*}/M_{\odot}<10^{10}$ & $11.72 \pm 0.03$ & $0.21 \pm 0.01$ & $0.95 \pm 0.09$ & $0.00 \pm 0.05$ & $0.55 \pm 0.05$ & $0.04 \pm 0.02$ \\
$10^{10}<M_{*}/M_{\odot}<10^{10.5}$ & $12.07 \pm 0.05$ & $0.20 \pm 0.01$ & $1.27 \pm 0.23$ & $0.32 \pm 0.15$ & $0.97 \pm 0.16$ & $0.016 \pm 0.015$ \\
$10^{10.5}<M_{*}/M_{\odot}<10^{11}$ & $12.33 \pm 0.03$ & $0.19 \pm 0.05$ & $1.5 \pm 0.6$ & $0.90 \pm 0.15$ & $0.28 \pm 0.52$ & $0.004 \pm 0.009$ \\
\hline
{\bf{SFR range}} & {\bf{$\log_{10}M_{\rm{min}}$}} & {\bf{$\sigma_{\log M}$}} & {\bf{$\alpha$}} & {\bf{$F_{c}^{A}$}} & {\bf{$F_{c}^{B}$}} & {\bf{$F_{s}$}} \\
\hline
$\rm{SFR}/M_{\odot}\rm{yr}^{-1}>10^{-1}$ & $11.43 \pm 0.04$ & $0.28 \pm 0.03$ & $0.70 \pm 0.08$ & $0.78 \pm 0.18$ & $0.06 \pm 0.05$ & $0.08 \pm 0.03$ \\
$\rm{SFR}/M_{\odot}\rm{yr}^{-1}>10^{-0.5}$ & $11.65 \pm 0.02$ & $0.22 \pm 0.01$ & $0.76 \pm 0.07$ & $0.72 \pm 0.08$ & $0.21 \pm 0.11$ & $0.05 \pm 0.02$ \\
$\rm{SFR}/M_{\odot}\rm{yr}^{-1}>10^{0}$ & $11.96 \pm 0.03$ & $0.21 \pm 0.04$ & $0.67 \pm 0.22$ & $0.50 \pm 0.15$ & $0.06 \pm 0.07$ & $0.04 \pm 0.04$ \\
$\rm{SFR}/M_{\odot}\rm{yr}^{-1}>10^{0.5}$ & $12.31 \pm 0.15$ & $0.21 \pm 0.23$ & $0.74 \pm 0.81$ & $0.09 \pm 0.21$ & $0.01 \pm 0.07$ & $0.018 \pm 0.06$ \\
\hline
$10^{-1}<\rm{SFR}/M_{\odot}\rm{yr}^{-1}<10^{-0.5}$ & $11.50 \pm 0.02$ & $0.20 \pm 0.01$ & $0.63 \pm 0.06$ & $0.00 \pm 0.00$ & $0.39 \pm 0.03$ & $0.057 \pm 0.018$ \\
$10^{-0.5}<\rm{SFR}/M_{\odot}\rm{yr}^{-1}<10^{0}$ & $11.58 \pm 0.04$ & $0.20 \pm 0.01$ & $1.11 \pm 0.14$ & $0.47 \pm 0.05$ & $0.09 \pm 0.08$ & $0.005 \pm 0.003$ \\
$10^{0}<\rm{SFR}/M_{\odot}\rm{yr}^{-1}<10^{0.5}$ & $12.00 \pm 0.09$ & $0.19 \pm 0.01$ & $1.5 \pm 0.5$ & $0.57 \pm 0.06$ & $0.25 \pm 0.29$ & $0.0006 \pm 0.0016$ \\
$10^{0.5}<\rm{SFR}/M_{\odot}\rm{yr}^{-1}<10^{1}$ & $12.30 \pm 0.16$ & $0.20 \pm 0.25$ & $0.60 \pm 0.13$ & $0.07 \pm 0.3$ & $0.01\pm 0.07$ & $0.03 \pm 0.09$ \\
\hline
\end{tabular}
\caption{Parameters of the halo occupation distribution model detailed in Section \ref{sec:func_forms}, fitted to EAGLE galaxies selected by different stellar mass and star-formation rate criteria. Figure \ref{fig:EAGLE_hods} shows that our chosen parametrization is flexible enough to provide good fits to HODs of very differently defined samples.}
\label{table:hod_params_fitted}
\end{center}
\end{table*}
The Halo Occupation Distribution describes the bias between galaxies and total mass by quantifying the average number of galaxies per dark matter halo as a function of halo mass. As described in the introduction, it is an important tool for linking the physics of galaxy evolution to the host halo environment, and frequently used derive host halo masses and satellite fractions from observed galaxy two-point functions for galaxies of different types. However, the reliability of this technique is highly dependent on the appropriate choice of an HOD parametrization. \\
\indent \cite{Kravtsov2003} proposed that the overall HOD can be parametrized by two simple terms. The first describes the probability that a dark matter halo of mass $M_{\rm{halo}}$ hosts a central galaxy above some stellar mass limit; this is well-approximated by a step function. Below some minimum halo mass, galaxies will not be found, since energy feedback from supernovae will simply expel baryons from very shallow potential wells, while above $M_{\rm{min}}$ all haloes host a galaxy. The second term describes the average number of satellite galaxies as a function of halo mass; empirically, this is well-fitted by a power law, for which a slope of unity appears to be appropriate for a wide range of simulated galaxy number densities and redshifts. Parametrizations of this form have been used fairly successfully for many years, for a variety of galaxy types and redshifts \citep[e.g.][]{Zehavi2005,Zheng2005,Zheng2007,Tinker2010a,Zehavi2011,Wake2011,Durkalec2015}. \\ 
\indent However, this simple parametrization becomes inappropriate when considering only sub-populations of galaxies, for example those within some stellar mass range, or above some star-formation rate. Here, it is clear that, although a halo may have a central galaxy, this galaxy may not satisfy the sample selection criteria. \cite{Geach2012} argued that a two-component HOD model, composed of a Gaussian distribution at low halo masses and a step function, was more appropriate for centrals in star-formation rate-limited samples, based on the output of {\small\it{GALFORM}} semi-analytic modelling \citep[e.g.][]{Bower2006}. \cite{Contreras2013} followed this with a detailed study of galaxies drawn from the Durham and Munich semianalytic models. The halo occupation of galaxies selected above a limiting cold gas mass or star-formation rate were better fitted by an asymmetric peak at low halo masses than by the traditional step function. \\
\indent Given the current availability of large samples of galaxies, increasingly samples can be split into stellar mass or star-formation rate bins too. It is unclear whether the HOD parametrizations adopted for limited samples are still appropriate. In \cite{Cochrane2017} we adopted a 6-parameter functional form for the HOD, based on the parametrization of \cite{Geach2012}, and showed that it did well in fitting observed two-point angular correlation functions of samples of galaxies binned by $\rm{H}\alpha$ luminosity, which broadly traces star-formation rate. Here, we briefly study the typical halo occupations of EAGLE galaxies, as a function of stellar mass and star-formation rate. The great advantage of EAGLE is that properties of the dark matter haloes and the galaxies within them are easily accessible, and so can provide the functional form of the HOD, for use in observational studies.\subsection{HOD functional forms from EAGLE}\label{sec:func_forms}
\begin{figure*} 
	\centering
	\includegraphics[scale=0.5]{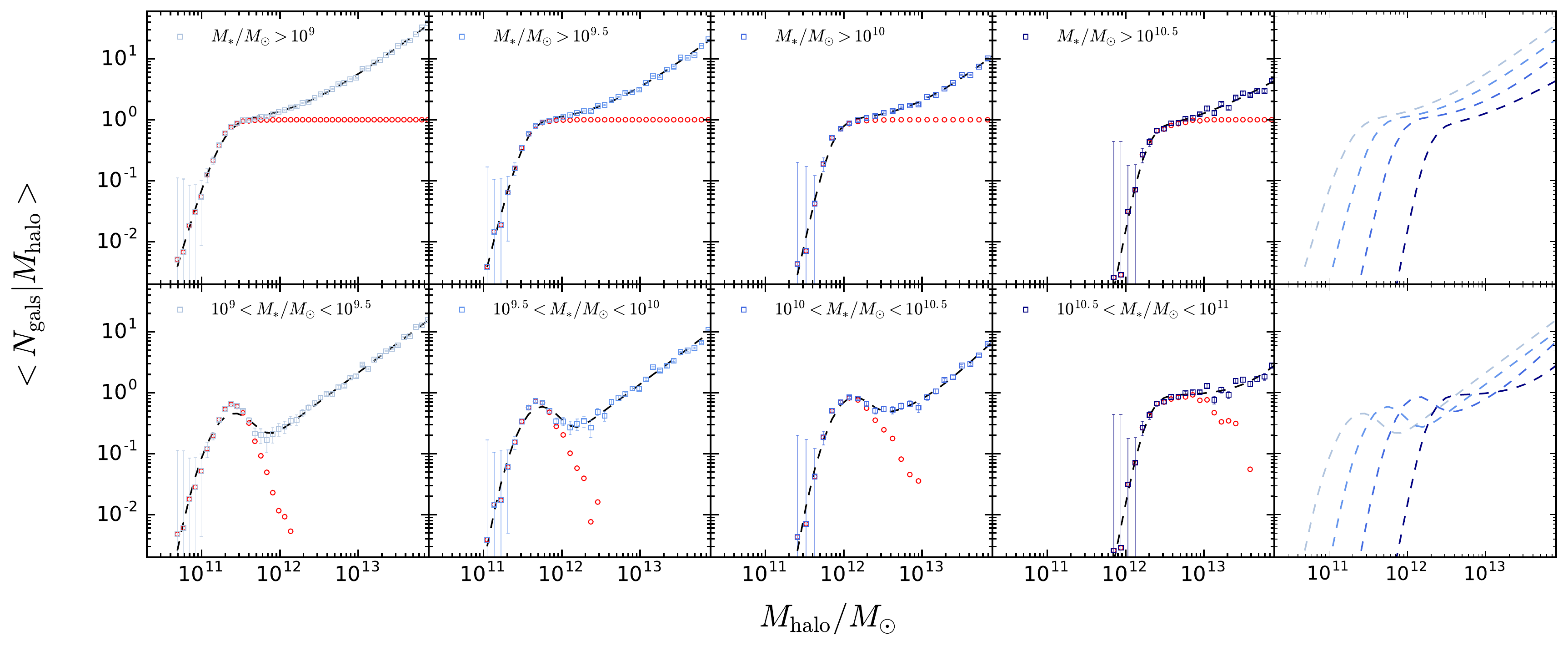}
	\includegraphics[scale=0.5]{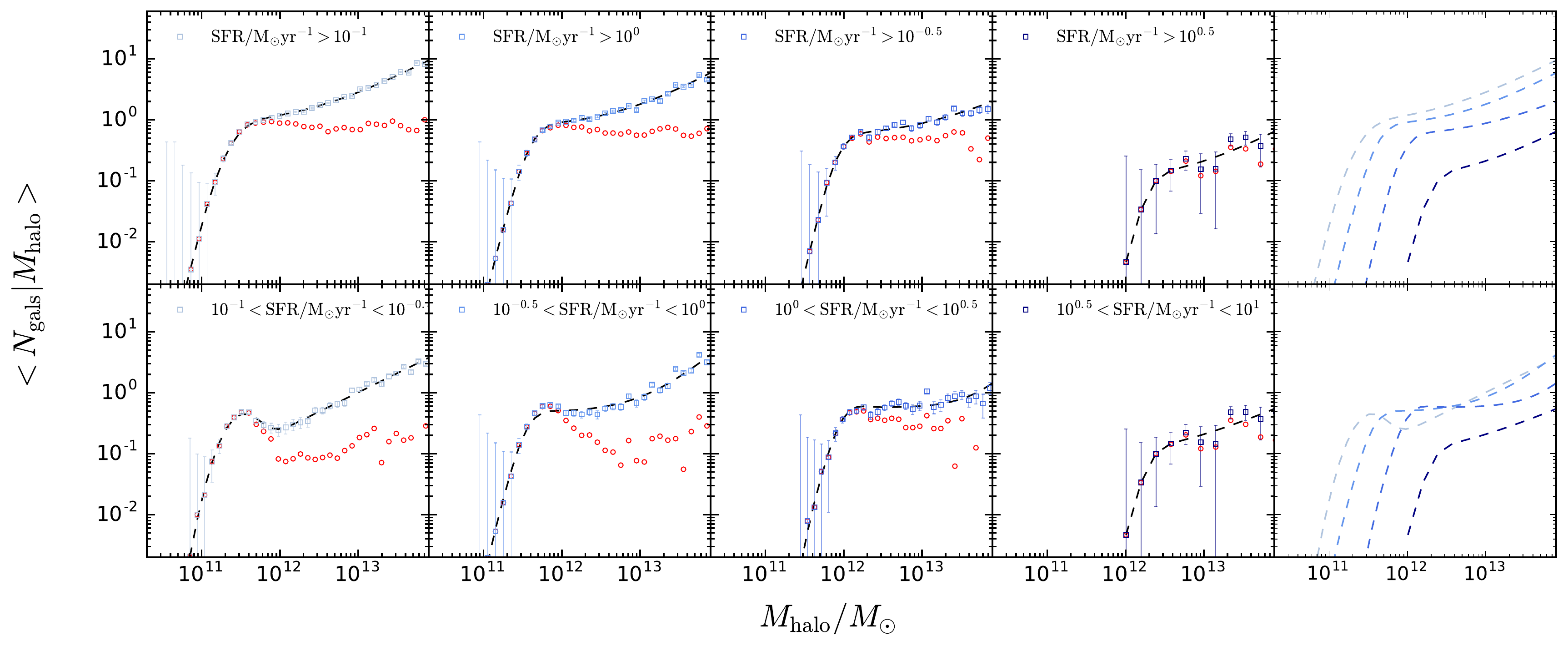}
	\caption{Halo Occupation Distributions constructed using EAGLE galaxies at $z=0.00$, with stellar mass and star-formation rate cuts (either limits, or binned ranges) applied. The blue squares show the whole galaxy population (centrals and satellites) and the red circles show only central galaxies. The dashed lines show the best-fitting HOD, given the parametrization presented in Section \ref{sec:func_forms}. It is encouraging that all samples (SFR and $M_{*}$ selected; binned and limited) can be well-matched using the same 6-parameter functional form. The best-fitting HODs are shown together in a separate panel (right) to show the differences between the samples more clearly. In general, more massive and more highly star-forming galaxies occupy more massive dark matter haloes. Parameters for all of these fits are provided in Table \ref{table:hod_params_fitted}.}
    \label{fig:EAGLE_hods}
\end{figure*}
Figure \ref{fig:EAGLE_hods} shows HODs for samples of EAGLE galaxies at $z=0.00$, given different mass and SFR cuts. The blue squares show the total (central \& satellite) occupancy, and red circles show only the central occupancy. For mass-limited samples (top row of Figure \ref{fig:EAGLE_hods}), the traditional smoothed step function appears a reasonable choice of parametrization. Occupancy of centrals flattens at unity, as expected. However, for mass-binned samples (second row of Figure \ref{fig:EAGLE_hods}), the red circles, which represent the HOD for central galaxies only, have a Gaussian-like form. This is very different to the canonical step-like function usually assumed in HOD fitting to mass-limited samples. Where samples are mass-incomplete, the central galaxy occupation does not rise and flatten at 1 at high halo masses, as for the mass-limited samples, because not all haloes contain a central galaxy within the chosen stellar mass range. \\
\indent For star-formation rate-limited samples (third row), at the lowest SFR limits the HOD is similar to those of the mass-limited samples. For higher star-formation rate cuts, the smoothed step-like function peaks below unity, since such samples do not include all the low star-formation rate galaxies that fall into a mass-selected sample. HODS are different again for star-formation rate-binned samples (bottom row). Here, although we see a peak in occupation at low halo masses, similar to the mass-binned samples, the HOD does not follow the Gaussian-like form above the peak. Instead, the occupation flattens at high halo masses, but at a value below unity. We thus urge caution in adopting standard forms of the HOD, and suggest that simulations such as EAGLE might be queried for specific classes of galaxies in order to obtain appropriate functional forms that may then be fitted to observed clustering measurements. \\
\indent Here, we present one functional form that appears to do well in fitting EAGLE galaxy HODs. We have adopted the flexible 6-parameter form used by \cite{Cochrane2017,Cochrane2017a} to fit observed galaxy clustering, which was based on the parametrization first presented by \cite{Geach2012}. The numbers of central and satellite galaxies are parametrized as:\\
\begin{equation}
\begin{split}
\langle N_{\rm{cen}}|\rm{M}\rangle = F_{c}^{B}(1-F_{c}^{A})\rm{exp}\Bigg[-\frac{\log(M/M_{min})^{2}}{2(\sigma_{\log M})^{2}}\Bigg] \\
+ \frac{1}{2}F_{c}^{A}\Bigg[1+\rm{erf}\Bigg(\frac{\log(M/M_{min})}{\sigma_{\log M}}\Bigg)\Bigg],
\end{split}
\end{equation}
\begin{equation}
\langle N_{\rm{sat}}|\rm{M}\rangle = F_{s}\Bigg[1+\rm{erf}\Bigg(\frac{\log(M/M_{\rm{min}})}{\sigma_{\log M}}\Bigg)\Bigg]\Bigg(\frac{M}{M_{\rm{min}}}\Bigg)^{\alpha},
\end{equation}
with the total number of galaxies given by: 
\begin{equation}
\langle N|M\rangle = \langle N_{\rm{cen}}|M\rangle + \langle N_{\rm{sat}}|M\rangle.
\end{equation}
The key parameters are:
\begin{itemize}[leftmargin=0.3cm]
  \item[--] $M_{\rm{min}}$: the minimum halo mass that hosts a galaxy. Note that this definition differs subtly to that used in work characterising mass-limited samples, since here $M_{\rm{min}}$ applies to both central and satellite galaxies.
  \item[--] $\sigma_{\log M}$: characterises the width of the transition to  $\langle N_{\rm{sat}}|M\rangle = F_{s} \Big(\frac{M}{M_{\rm{min}}}\Big)^{\alpha}$ around $M_{\rm{min}}$.
  \item[--] $\alpha$: the slope of the power-law for $\langle N_{\rm{sat}}|M\rangle$ in haloes with $M>M_{\rm{min}}$.
  \item[--] $F_{c}^{A,B}$: normalisation factors, in range [0,1].
  \item[--] $F_{s}$: the mean number of satellite galaxies per halo, at $M = M_{\rm{min}}$
  \end{itemize}
Using this parametrization, we denote the best-fitting HODs for each subsample in Figure \ref{fig:EAGLE_hods} by a dashed black line, and provide details of the parameter estimates in Table \ref{table:hod_params_fitted}. From the successful fits shown, it is clear that this parametrization is appropriate for a wide range of stellar mass and star-formation rate selected samples (both binned and limited). Where central galaxies occupy only the lower halo masses, there is a clear Gaussian component to the HOD but no step-function-like occupation at higher halo masses. This is the case for the lowest two stellar mass bins ($10^{9}<M_{*}/M_{\odot}<10^{9.5}$ and $10^{9.5}<M_{*}/M_{\odot}<10^{10}$). Here, $F_{c}^{A}$, the step-function normalisation, becomes vanishingly small and $F_{c}^{B}$, which determines the contribution from the low-halo mass Gaussian component, dominates. For the stellar mass-limited samples, the contribution from $F_{c}^{A}$ is close to unity, and that of $F_{c}^{B}$ generally consistent with zero. \\
\indent While the slope of the power-law occupancy of satellite galaxies, $\alpha$, is well-approximated by unity for the mass-limited and mass-binned samples, this appears less suitable for the star-formation rate-selected samples. For these, our fits favour a lower $\alpha$, indicative of satellite quenching in high mass haloes, which removes galaxies from samples selected by star-formation rate.
\section{Full details of principal components at each redshift}\label{sec:appendix1}
\begin{table*}
\begin{center}
\begin{tabular}{l|c|c|c|c|c|c}
{\bf{Redshift}} & {\bf{PC1}} & {\bf{PC1 Var}} & {\bf{PC2}} & {\bf{PC2 Var}} & {\bf{PC3}} & {\bf{PC3 Var}} \\
\hline
\multicolumn{6}{l}{Centrals, $10^{10}M_{\odot}<M_{\rm{halo}}<10^{12}M_{\odot}$} \\
\hline
0.00 & [0.59, 0.60, 0.54] & 78.6\% & [0.46, 0.31, -0.83] & 14.4\% & [0.67, -0.74, 0.09] & 7.0\% \\
0.10 & [0.59, 0.59, 0.55] & 79.6\% & [0.44, 0.33, -0.83] & 13.6\% & [0.68, -0.73, 0.07] & 6.8\% \\
0.18 & [0.59, 0.59, 0.55] & 80.1\% & [0.42, 0.36, -0.83] & 13.2\% & [0.69, -0.72, 0.04] & 6.7\%  \\
0.27 & [0.59, 0.59, 0.55] & 81.5\% & [0.43, 0.35, -0.83] & 12.1\% & [0.68, -0.73, 0.05] & 6.4\% \\
0.37 & [0.58, 0.59, 0.56] & 82.4\% & [0.45, 0.33, -0.83] & 11.0\% & [0.67, -0.74, 0.07] & 6.6\% \\
0.50 & [0.58, 0.59, 0.56] & 83.5\% & [0.44, 0.36, -0.82] & 10.0\% & [0.69, -0.73, 0.05] & 6.5\% \\
0.62 & [0.58, 0.58, 0.57] & 84.6\% & [0.51, 0.29, -0.81] & 9.0\% & [0.64, -0.76, 0.13] & 6.5\% \\
0.74 & [0.58, 0.59, 0.57] & 84.7\% & [0.59, 0.18, -0.79] & 8.9\% & [0.56, -0.79, 0.24] & 6.4\% \\
0.87 & [0.58, 0.58, 0.57] & 85.1\% & [0.60, 0.17, -0.78] & 8.4\% & [0.55, -0.79, 0.25] & 6.6\% \\
1.00 & [0.57, 0.58, 0.57] & 85.3\% & [0.68, 0.04, -0.73] & 8.3\% & [0.45, -0.81, 0.38] & 6.5\% \\
1.26 & [0.57, 0.58, 0.58] & 85.4\% & [0.76, -0.12, -0.63] & 8.3\% & [0.30, -0.80, 0.52] & 6.3\% \\
1.49 & [0.57, 0.58, 0.58] & 84.8\% & [0.78, -0.17, -0.60] & 8.7\% & [0.25, -0.79, 0.56] & 6.5\% \\
1.74 & [0.57, 0.59, 0.58] & 84.0\% & [0.78, -0.16, -0.61] & 9.2\% & [0.26, -0.79, 0.55] & 6.8\% \\
2.01 & [0.57, 0.58, 0.58] & 84.9\% & [0.81, -0.29, -0.50] & 8.9\% & [0.12, -0.76, 0.64] & 6.2\% \\
2.24 & [0.57, 0.58, 0.58] & 84.4\% & [0.82, -0.31, -0.49] & 9.2\% & [0.11, -0.75, 0.65] & 6.4\% \\
2.48 & [0.57, 0.59, 0.57] & 82.6\% & [0.73, -0.05, -0.68] & 10.0\% & [0.37, -0.81, 0.46] & 7.4\% \\
3.02 & [0.58, 0.59, 0.57] & 79.6\% & [0.57, 0.20, -0.79] & 11.8\% & [0.58, -0.78, 0.22] & 8.6\% \\
3.53 & [0.58, 0.59, 0.57] & 78.6\% & [0.57, 0.20, -0.79] & 12.1\% & [0.58, -0.78, 0.22] & 9.3\% \\
3.98 & [0.56, 0.59, 0.58] & 78.4\% & [0.80, -0.24, -0.55] & 12.4\% & [0.18, -0.77, 0.61] & 9.2\% \\
\hline
\multicolumn{6}{l}{Satellites, $10^{10}M_{\odot}<M_{\rm{halo}}<10^{12}M_{\odot}$} \\
\hline
0.00 & [0.53, 0.61, 0.59] & 58.1\% & [0.84, -0.28, -0.46] & 23.7\% & [0.11, -0.74, 0.66] & 18.2\% \\
0.10 & [0.50, 0.62, 0.60] & 58.2\% & [0.86, -0.27, -0.44] & 25.1\% & [0.11, -0.74, 0.67] & 16.7\% \\
0.18 & [0.48, 0.62, 0.61] & 56.8\% & [0.87, -0.30, -0.38] & 26.3\% & [0.05, -0.72, 0.69] & 16.9\% \\
0.27 & [0.49, 0.63, 0.60] & 58.5\%  & [0.86, -0.24, -0.45] & 25.7\% & [0.14, -0.74, 0.66] & 15.8\% \\
0.37 & [0.50, 0.63, 0.60] & 60.4\% & [0.86, -0.24, -0.45] & 24.8\% & [0.14, -0.74, 0.66] & 14.8\% \\
0.50 & [0.46, 0.64, 0.61] & 59.9\% & [0.88, -0.23, -0.42] & 26.3\% & [0.12, -0.73, 0.67] & 13.9\% \\
0.62 & [0.49, 0.61, 0.62] & 63.6\% & [0.87, -0.37, -0.32] & 23.9\% & [-0.04, -0.70, 0.72] & 12.4\% \\
0.74 & [0.47, 0.63, 0.62] & 62.8\% & [0.88, -0.27, -0.39] & 25.3\% & [0.08, -0.73, 0.68] & 11.9\% \\
0.87 & [0.46, 0.63, 0.63] & 63.5\% & [0.89, -0.31, -0.33] & 25.4\% & [0.02, -0.71, 0.70] & 11.1\%\\
1.00 & [0.48, 0.62, 0.62] & 66.1\% & [0.88, -0.34, -0.33] & 23.7\% & [-0.01, -0.70, 0.71] & 10.2\% \\
1.26 & [0.48, 0.62, 0.63] & 64.7\% & [0.88, -0.37, -0.30] & 24.0\% & [-0.05, -0.69, 0.72] & 11.3\% \\
1.49 & [0.47, 0.62, 0.63] & 67.4\% & [0.88, -0.36, -0.31] & 23.7\% & [-0.04, -0.70, 0.72] & 8.9\% \\
1.74 & [0.43, 0.63, 0.64] & 64.7\% & [0.90, -0.33, -0.28] & 26.2\% & [-0.03, -0.70, 0.71] & 9.1\% \\
2.01 & [0.45, 0.63, 0.63] & 65.2\% & [0.89, -0.30, -0.33] & 25.4\% & [-0.02, -0.71, 0.70] & 9.4\% \\
2.24 & [0.41, 0.64, 0.65] & 63.5\% & [0.91, -0.30, -0.27] & 27.4\% & [-0.02, -0.70, 0.71] & 9.1\% \\
2.48 & [0.48, 0.63, 0.62] & 60.0\% & [0.88, -0.30, -0.38] & 25.4\% & [0.05, -0.72, 0.69] & 14.6\% \\
3.02 & [0.50, 0.61, 0.61] & 68.5\% & [0.86, -0.37, -0.34] & 21.4\% & [-0.02, -0.70, 0.71] & 10.1\% \\
3.53 & [0.49, 0.64, 0.59] & 61.9\% & [0.84, -0.18, -0.51] & 24.9\% & [0.22, -0.75, 0.62] & 13.1\% \\
3.98 & [0.57, 0.58, 0.58] & 57.9\% & [0.77, -0.15, -0.61] & 21.4\% & [0.27, -0.80, 0.54] & 20.7\% \\
\end{tabular}
\caption{Principal components of $M_{*}>10^{9}M_{\odot}$, $\rm{SFR}>0M_{\odot}\rm{yr}^{-1}$, central and satellite EAGLE galaxies in low mass haloes ($10^{10}M_{\odot}<M_{\rm{halo}}<10^{12}M_{\odot}$) at each redshift. Vectors have ordering $[\log_{10}M_{\rm{halo}}/M_{\odot},\,\, \log_{10}M_{*}/M_{\odot},\,\, \log_{10}\rm{SFR}/M_{\odot}\rm{yr}^{-1}]$.}
\label{table:pca_table_z}
\end{center}
\end{table*}

\begin{table*}
\begin{center}
\begin{tabular}{l|c|c|c|c|c|c}
{\bf{Redshift}} & {\bf{PC1}} & {\bf{PC1 Var}} & {\bf{PC2}} & {\bf{PC2 Var}} & {\bf{PC3}} & {\bf{PC3 Var}} \\
\hline
\multicolumn{6}{l}{Centrals, $10^{12}M_{\odot}<M_{\rm{halo}}<10^{14}M_{\odot}$} \\
\hline
0.00 & [0.71, 0.71, 0.04] & 61.1\% & [-0.07, 0.01, 1.00] & 33.4\% & [0.71, -0.71, 0.06] & 5.5\% \\
0.10 & [0.70, 0.71, 0.10] & 60.8\% &  [-0.10, -0.03, 0.99] & 33.1\% & [0.70, -0.71, 0.05] & 6.1\% \\
0.18 & [0.70, 0.70, 0.13] & 61.1\% & [-0.13, -0.06, 0.99] & 32.9\% & [0.70, -0.71, 0.05] & 6.0\% \\
0.27 & [0.70, 0.70, 0.15] & 60.8\% & [-0.15, -0.07, 0.99] & 32.8\% & [0.70, -0.71, 0.06] & 6.4\% \\
0.37 & [0.69, 0.70, 0.20] & 61.2\% & [-0.19, -0.09, 0.98] & 32.3\% & [0.70, -0.71, 0.07] & 6.5\% \\
0.50 & [0.69, 0.70, 0.17] & 60.1\% & [-0.18, -0.07, 0.98] & 32.7\% & [0.70, -0.71, 0.08] & 7.2\% \\
0.62 & [0.68, 0.69, 0.24] & 60.7\% & [-0.20, -0.14, 0.97] & 31.7\% & [0.70, -0.71, 0.04] & 7.6\% \\
0.74 & [0.68, 0.69, 0.26] & 60.4\% & [-0.21, -0.16, 0.96] & 31.4\% & [0.70, -0.71, 0.03] & 8.2\% \\
0.87 & [0.68, 0.69, 0.23] & 59.2\% & [-0.20, -0.13, 0.97] & 31.9\% & [0.70, -0.71, 0.05] & 8.8\% \\
1.00 & [0.67, 0.69, 0.28] & 59.7\% & [-0.25, -0.15, 0.96] & 31.2\% & [0.70, -0.71, 0.07] & 9.0\% \\
1.26 & [0.66, 0.67, 0.34] & 60.3\% & [-0.29, -0.19, 0.94] & 29.9\% & [0.69, -0.72, 0.07] & 9.8\% \\
1.49 & [0.65, 0.67, 0.34] & 60.3\% & [-0.31, -0.17, 0.93] & 30.0\% & [0.69, -0.72, 0.10] & 9.7\% \\
1.74 & [0.65, 0.67, 0.36] & 60.5\% & [-0.30, -0.21, 0.93] & 29.5\% & [0.70, -0.72, 0.06] & 10.0\% \\
2.01 & [0.65, 0.66, 0.38] & 60.8\% & [-0.30, -0.23, 0.93] & 28.9\% & [0.70, -0.71, 0.04] & 10.4\% \\
2.24 & [0.65, 0.66, 0.39] & 60.3\% & [-0.31, -0.24, 0.92] & 28.7\% & [0.70, -0.72, 0.05] & 11.0\% \\
2.48 & [0.64, 0.65, 0.42] & 61.3\% & [-0.31, -0.28, 0.91] & 27.4\% & [0.70, -0.71, 0.02] & 11.3\% \\
3.02 & [0.60, 0.65, 0.46] & 59.6\% & [-0.47, -0.18, 0.86] & 26.7\% & [0.64, -0.74, 0.19] & 13.6\% \\
3.53 & [0.59, 0.65, 0.48] & 56.6\% & [-0.53, -0.13, 0.84] & 27.6\% & [0.61, -0.75, 0.27] & 15.8\% \\
3.98 & [0.59, 0.62, 0.52] & 59.1\% & [-0.48, -0.24, 0.84] & 24.2\% & [0.65, -0.75, 0.15] & 16.7\% \\
\hline
\multicolumn{6}{l}{Satellites, $10^{12}M_{\odot}<M_{\rm{halo}}<10^{14}M_{\odot}$} \\
\hline
0.00 & [0.39, 0.70, 0.60] & 54.9\% & [0.84, -0.02, -0.54] & 32.6\% & [0.37, -0.71, 0.60] & 12.5\% \\
0.10 & [0.33, 0.71, 0.62] & 54.2\% & [0.87, 0.02, -0.49] & 34.0\% & [0.36, -0.70, 0.61] & 11.8\% \\
0.18 & [0.34, 0.71, 0.61] & 54.4\% & [0.87, 0.01, -0.50] & 33.8\% & [0.36, -0.70, 0.61] & 11.7\% \\
0.27 & [0.34, 0.71, 0.62] & 53.6\% & [0.87, 0.01, -0.50] & 33.8\% & [0.36, -0.70, 0.61] & 12.6\% \\
0.37 & [0.30, 0.72, 0.63] & 53.8\% & [0.89, 0.03, -0.46] & 34.3\% & [0.35, -0.70, 0.62] & 11.9\% \\
0.50 & [0.29, 0.72, 0.63] & 53.9\% & [0.89, 0.05, -0.46] & 34.7\% & [0.36, -0.69, 0.62] & 11.4\% \\
0.62 & [0.29, 0.71, 0.64] & 54.3\% & [0.89, 0.04, -0.44] & 34.3\% & [0.34, -0.70, 0.63] & 11.4\% \\
0.74 & [0.25, 0.72, 0.65] & 54.3\% & [0.91, 0.06, -0.41] & 34.7\% & [0.33, -0.69, 0.64] & 11.0\% \\
0.87 & [0.22, 0.72, 0.66] & 53.8\% & [0.92, 0.08, -0.39] & 35.2\% & [0.34, -0.69, 0.64] & 11.0\% \\
1.00 & [0.21, 0.72, 0.66] & 55.0\% & [0.92, 0.07, -0.38] & 34.9\% & [0.32, -0.69, 0.65] & 10.1\% \\
1.26 & [0.26, 0.71, 0.66] & 56.2\% & [0.93, 0.01, -0.38] & 33.6\% & [0.28, -0.70, 0.65] & 10.2\% \\
1.49 & [0.27, 0.70, 0.66] & 56.6\% & [0.93, -0.02, -0.36] & 32.8\% & [0.24, -0.71, 0.66] & 10.6\% \\
1.74 & [0.21, 0.71, 0.67] & 55.3\% & [0.95, 0.02, -0.33] & 33.7\% & [0.25, -0.70, 0.67] & 10.9\% \\
2.01 & [0.27, 0.71, 0.66] & 54.6\% & [0.93, -0.01, -0.37] & 33.2\% & [0.25, -0.71, 0.66] & 12.2\% \\
2.24 & [0.21, 0.71, 0.67] & 52.2\% & [0.93, 0.05, -0.35] & 34.2\% & [0.28, -0.70, 0.66] & 13.6\% \\
2.48 & [0.18, 0.72, 0.68] & 52.1\% & [0.94, 0.08, -0.33] & 34.6\% & [0.29, -0.69, 0.66] & 13.3\% \\
3.02 & [0.41, 0.67, 0.61] & 55.9\% & [0.88, -0.11, -0.47] & 29.9\% & [0.25, -0.73, 0.63] & 14.3\% \\
3.53 & [0.43, 0.70, 0.57] & 51.5\% & [0.81, -0.02, -0.59] & 32.5\% & [0.40, -0.72, 0.57] & 16.0\% \\
3.98 & [0.46, 0.66, 0.59] & 52.0\% & [0.84, -0.11, -0.54] & 29.3\% & [0.29, -0.74, 0.60] & 18.7\% \\
\end{tabular}
\caption{Principal components of $M_{*}>10^{9}M_{\odot}$, $\rm{SFR}>0M_{\odot}\rm{yr}^{-1}$, central and satellite EAGLE galaxies in high mass haloes ($10^{12}M_{\odot}<M_{\rm{halo}}<10^{14}M_{\odot}$) at each redshift. Vectors have ordering $[\log_{10}M_{\rm{halo}}/M_{\odot},\,\, \log_{10}M_{*}/M_{\odot},\,\, \log_{10}\rm{SFR}/M_{\odot}\rm{yr}^{-1}]$.}
\label{table:pca_table_z}
\end{center}
\end{table*}


\label{lastpage}
\end{document}